\documentclass[letterpaper,5p]{elsarticle}

\usepackage{siunitx}
\usepackage{dblfloatfix}
\usepackage{pgfplots,tikz}
\pgfplotsset{compat=1.14}
\usetikzlibrary{datavisualization, datavisualization.formats.functions}
\usepackage{hyperref}
\usepackage{footnotebackref}

\usepackage{cleveref}

\journal{arXiv}
\DeclareSIUnit\KWH{kWh}
\newcommand\kpWavelength{\SI{1.06}{\micro\meter}}
\newcommand\kpEmittance{0.01}
\newcommand\kpAbsorptance{$10^{-8}$}
\newcommand\kpMaxTemp{\SI{625}{\kelvin}}
\newcommand\kpZeroTwoCCapex{\$8.0B}
\newcommand\kpZeroTwoCLasers{\$2.0B}
\newcommand\kpZeroTwoCOptics{\$2.8B}
\newcommand\kpZeroTwoCStorage{\$3.1B}
\newcommand\kpZeroTwoCRelKE{\SI{1.9}{GWh}}
\newcommand\kpZeroTwoCRelKETJ{\SI{6.7}{TJ}}
\newcommand\kpZeroTwoCStored{\SI{63}{GWh}}
\newcommand\kpZeroTwoCEnergy{\$6M}
\newcommand\kpZeroTwoCPowerMax{\SI{200}{\giga\watt}}
\newcommand\kpZeroTwoCSysEfficiency{2.9\%}
\newcommand\kpZeroTwoCSailDiameter{\SI{4.1}{\meter}} 
\newcommand\kpZeroTwoCBeamerDiameter{\SI{2.7}{\kilo\meter}}
\newcommand\kpZeroTwoCSailMass{\SI{3.6}{\gram}}
\newcommand\kpZeroTwoCDurationPulseSec{\SI{480}{\second}}
\newcommand\kpZeroTwoCDurationPulseMin{\SI{8}{\minute}}
\newcommand\kpZeroTwoCDurationAccnSec{\SI{550}{\second}}
\newcommand\kpZeroTwoCDurationAccnMin{\SI{9}{\minute}}
\newcommand\kpZeroTwoCPhotonPressure{\SI{40}{\pascal}}
\newcommand\kpZeroTwoCPhotonForce{\SI{520}{\newton}}
\newcommand\kpZeroTwoCAccnInit{\SI{14900}{g's}}
\newcommand\kpZeroTwoCAccnFin{\SI{2500}{g's}}
\newcommand\kpZeroTwoCDispFinAu{\SI{0.13}{au}}
\newcommand\kpZeroTwoCDispFinLs{\SI{67}{ls}}
\newcommand\kpZeroTwoCFluxBeamerMax{\SI{37}{\kilo\watt\per\meter\squared}}
\newcommand\kpZeroTwoCFluxSailTLimited{\SI{8.7}{\giga\watt\per\meter\squared}}
\newcommand\kpZZOCCapex{\$517M}
\newcommand\kpZZOCLasers{\$285M}
\newcommand\kpZZOCOptics{\$224M}
\newcommand\kpZZOCStorage{\$8M}
\newcommand\kpZZOCRelKE{\SI{8}{kWh}}
\newcommand\kpZZOCRelKEMJ{\SI{30}{MJ}}
\newcommand\kpZZOCRelKEspecific{\SI{4}{\tera\joule\per\kilo\gram}}
\newcommand\kpZZOCStored{\SI{78}{MWh}}
\newcommand\kpZZOCEnergy{\$8k}
\newcommand\kpZZOCPowerMax{\SI{285}{\mega\watt}}
\newcommand\kpZZOCSysEfficiency{0.01\%}
\newcommand\kpZZOCSailDiameter{\SI{19}{\centi\meter}} 
\newcommand\kpZZOCBeamerDiameter{\SI{169}{\meter}}
\newcommand\kpZZOCSailMass{\SI{6.6}{\milli\gram}}
\newcommand\kpZZOCDurationPulseSec{\SI{346}{\second}}
\newcommand\kpZZOCDurationPulseMin{\SI{6}{\minute}}
\newcommand\kpZZOCDurationAccnSec{\SI{349}{\second}}
\newcommand\kpZZOCDurationAccnMin{\SI{6}{\minute}}
\newcommand\kpZZOCPhotonPressure{\SI{23}{\pascal}}
\newcommand\kpZZOCPhotonForce{\SI{0.64}{\newton}}
\newcommand\kpZZOCAccnInit{\SI{10000}{g's}}
\newcommand\kpZZOCAccnFin{\SI{7}{g's}}
\newcommand\kpZZOCDispFinAu{\SI{0.007}{au}}
\newcommand\kpZZOCDispFinLs{\SI{3.3}{ls}}
\newcommand\kpZZOCFluxBeamerMax{\SI{13}{\kilo\watt\per\meter\squared}}
\newcommand\kpZZOCFluxSailTLimited{\SI{8.7}{\giga\watt\per\meter\squared}}

\newcommand\kpTunnelInputOptics{\$1M/m\textsuperscript{2}}
\newcommand\kpTunnelInputTunnel{\$10k/m}
\newcommand\kpTunnelAccnCCapex{\$5M}
\newcommand\kpTunnelAccnTLen{\SI{0.4}{\kilo\meter}}

\newcommand\kpTunnelPwr{\SI{22}{\kilo\watt}}
\newcommand\kpTunnelOpticsDia{\SI{0.6}{\meter}}

\newcommand\kpTunnelTwoAccnCCapex{\$1.3B}
\newcommand\kpTunnelTwoAccnTLen{\SI{28}{\kilo\meter}}

\newcommand\kpTunnelTwoPwr{\SI{4.6}{\mega\watt}}
\newcommand\kpTunnelTwoOpticsDia{\SI{3.1}{\meter}}

\begin{document}

\begin{frontmatter}

\title{The Breakthrough Starshot System Model}
\author{Kevin L. G. Parkin\corref{kpfn}}
\address{Parkin Research LLC, 2261 Market Street \#221, San Francisco, USA}
\ead{kevin@parkinresearch.com}
\cortext[kpfn]{Systems Director, Breakthrough Starshot}

\begin{abstract}

Breakthrough Starshot is an initiative to prove ultra-fast light-driven nanocrafts, and lay the foundations for a first launch to Alpha Centauri within the next generation. Along the way, the project could generate important supplementary benefits to solar system exploration. A number of hard engineering challenges remain to be solved before these missions can become a reality.

A system model has been formulated as part of the Starshot systems engineering work. This paper presents the model and describes how it computes cost-optimal point designs. Three point designs are computed: A \SI{0.2}{c} mission to Alpha Centauri, a \SI{0.01}{c} solar system precursor mission, and a ground-based test facility based on a vacuum tunnel. All assume that the photon pressure from a \kpWavelength\ wavelength beam accelerates a circular dielectric sail. The \SI{0.2}{c} point design assumes \SI[per-mode=symbol]{0.01}[\$]{\per\watt} lasers, \SI[per-mode=symbol]{500}[\$]{\per\meter\squared} optics, and \SI[per-mode=symbol]{50}[\$]{\per\KWH} energy storage to achieve \kpZeroTwoCCapex\ capital cost for the ground-based beam director. In contrast, the energy needed to accelerate each sail costs \kpZeroTwoCEnergy. Beam director capital cost is minimized by a \kpZeroTwoCSailDiameter\ diameter sail that is accelerated for \kpZeroTwoCDurationAccnMin. The \SI{0.01}{c} point design assumes \SI[per-mode=symbol]{1}[\$]{\per\watt} lasers, \$10k/m\textsuperscript{2} optics, and \SI[per-mode=symbol]{100}[\$]{\per\KWH} energy storage to achieve \kpZZOCCapex\ capital cost for the beam director and \kpZZOCEnergy\ energy cost to accelerate each \kpZZOCSailDiameter\ diameter sail. The ground-based test facility assumes \SI[per-mode=symbol]{100}[\$]{\per\watt} lasers, \kpTunnelInputOptics\ optics, \SI[per-mode=symbol]{500}[\$]{\per\KWH} energy storage, and \kpTunnelInputTunnel\ vacuum tunnel. To reach \SI{20}{\kilo\meter\per\second}, fast enough to escape the solar system from Earth, takes \kpTunnelAccnTLen\ of vacuum tunnel, \kpTunnelPwr\ of lasers, and a \kpTunnelOpticsDia\ diameter telescope, all of which costs \kpTunnelAccnCCapex.

The system model predicts that, ultimately, Starshot can scale to propel probes faster than \SI{0.9}{c}.

\end{abstract}


\end{frontmatter}

\section{Objectives}
\paragraph{Breakthrough Starshot}
Send 1 gram of scientific instrumentation to the Centauri System to study it.  Image its planets, look for life and transmit the results to Earth.  Do so by using a beam emitted from the Earth to accelerate a sail carrying the instrumentation to \SI{0.2}{c}.  The capital cost of the equipment shall be less than \$10B.
\paragraph{Systems Engineering Work}
Ensure that Starshot engineering activities amount to a mission that fulfills the Breakthrough Starshot objectives.
\paragraph{System Model}
Replace physical experiments with simulations in cases where it saves time and money to do so. Verify Breakthrough Starshot feasibility and estimate performance. Design, optimize, trade-off, and analyze alternatives. Generate and quantify requirements. Model the impact of changes.
\section{Context}
\begin{figure*}[b]
\centering
\includegraphics[width=0.9\textwidth]{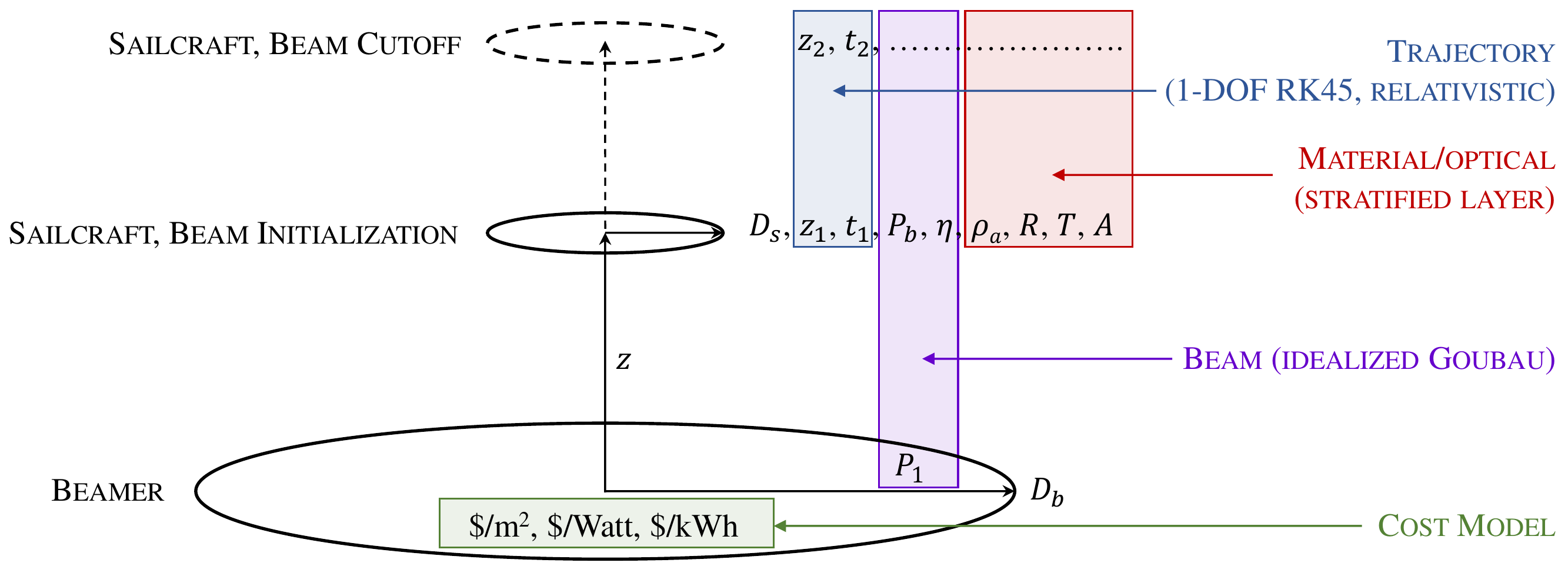}
\caption{System model}
\label{fig:systemmodel}
\end{figure*}
The first operational laser was built by Maiman of Hughes Research Labs in 1960 \cite{maiman1960stimulated}. Only two years later, Robert Forward, also of Hughes Research Labs, proposed to use the photon pressure from lasers to accelerate sails to speeds that bridge interstellar distances \cite{forward1962pluto}. In his 1984 paper \cite{forward1984roundtrip}, Forward suggests coherently combining many lasers into a phased array to reach high enough power and large enough primary optic sizes. He correctly identifies the key figure of merit as the sail acceleration produced per unit laser power. This figure of merit favors a sail material that is thin, light, reflective, and has low beam absorptance. However, Forward considers only metallic sails, which are absorption-limited to fluxes of less than \SI{10}{\kilo\watt\per\meter\squared}. The resulting \SI{3.6}{\kilo\meter} diameter, \SI{1}{\tonne} sail accelerates to \SI{0.11}{c} over a period of three years using a \SI{65}{\giga\watt} laser whose primary optic is \SI{1000}{\kilo\meter} in diameter.

In his 1986 paper \cite{forward1986laser}, Forward makes key conceptual progress by suggesting that sails be made from multilayer dielectrics instead of metals. Dielectrics have more than five orders of magnitude lower absorption than metals because their electrons are bound electrons that cannot absorb photons except at discrete wavelengths.  In comparison, metals absorb energy from an incident beam via the mechanism of Joule heating. The metal's softening temperature limits the irradiance on the sail, in turn limiting its acceleration.  Switching from metals to dielectrics profoundly increases the irradiance and acceleration of the sail, which in turn decreases the timescale, beam size, and cost.  Using eight quarter-wave thick layers of diamond, Forward's \SI{6.4}{\centi\meter} diameter, \SI{4.5}{\milli\gram} sail accelerates to \SI{0.015}{c} over a period of \SI{2.6}{\second} using a \SI{1}{\giga\watt} laser whose primary optic is \SI{100}{\meter} in diameter.

The dielectric sail concept was improved in 1989 when Landis pointed out that a single quarter-wave thick dielectric film accelerates faster than a multilayer film despite having lower reflectance \cite{landis1989optics}. In unpublished work, Landis went on to find that because of the way that reflectance scales with thickness, the true optimum sail thickness is somewhat less than a quarter wavelength. His result is confirmed later in this paper.

In work leading directly to the formation of Breakthrough Starshot in early 2016, a Harvard team led by Loeb \cite{guillochon2015seti} used a simple spreadsheet model to verify the feasibility and performance of a beam-driven sail and to produce point designs. A UCSB team led by Lubin \cite{lubin2016roadmap} developed a comprehensive analytical model that sought to describe the leading order system behavior. To use closed-form equations and avoid numerical trajectory integration, both the Harvard and UCSB models used simplifying approximations including top-hat beams.  To the extent that cost was minimized, it was minimized by manual experimentation with the model's input values.

The present system model was formulated in March 2016 and set out to verify the assumptions and results of the earlier models.  Key questions at this stage were: Are the earlier models correct? What diameter is the sailcraft\footnote{sail including payload} and beamer\footnote{beam director}? How much smaller/cheaper can the beamer be if the sailcraft continues to accelerate until the beam becomes too weak? How cheap must the lasers/microwaves be for the beamer to cost less than \$10B?

\section{Formulation}
The system model is formulated around the propagation of a beam from a ground-level beamer to a sailcraft in space above it, as shown in \cref{fig:systemmodel}.  The sailcraft begins at a given initial displacement above the beamer.  This displacement, in combination with the beamer diameter, is used by a beam propagation model to determine the fraction of transmitted power that reaches the sailcraft.  A material/optical model calculates how much of the power that is incident on the sailcraft is reflected or absorbed.  A relativistic equation of motion then translates this power into an acceleration.  The equation of motion is numerically integrated forward in time until the sailcraft reaches its desired cruise velocity.  The last photons arriving at the sailcraft are traced back in space and time to determine when beam cutoff occurs at the beamer. The system parameters of beamer diameter, beamer power, and sailcraft diameter are optimized to ensure that the sailcraft actually reaches cruise velocity and does so using a minimum-cost beamer.  This cost optimization reduces the dimensionality of the model because the beamer diameter, beamer power, and sailcraft diameter are no longer inputs.

\subsection{Goubau Beam}

In the general case, beam propagation from a ground-level beamer to a space-based sailcraft involves models representing a phased array, its elements, and transatmospheric propagation. This system model simplifies the beamer to an effective primary optic that transmits an idealized beam. The system model makes no representations about the technologies used in the array elements, the element sizes, or how the phased array as a whole is implemented.  Nor does the system model make detailed estimates of atmospheric attenuation.  Instead, all these factors are rolled into a user-provided value for stored to transmitted power efficiency, and a user-provided value for atmospheric attenuation.  The system model does calculate the idealized beam transfer efficiency, which varies the most of these three efficiency factors.

Unlike Gaussian beams and top-hat beams, Goubau beams \cite{goubau1961guided,goubau19683} describe near-optimal energy transfer between \textit{finite} optics \cite{hansen2005universal}.  For this reason, Goubau beams are used in the context of wireless power transfer \cite{brown1992beamed}.  When high transfer efficiency is needed, the beam profile resembles a Gaussian, and when low transfer efficiency can be tolerated, the profile resembles a top-hat.  At intermediate efficiencies, the beam resembles a truncated Gaussian. But which transfer efficiency minimizes the beamer cost? Assumptions about the answer to this question are often wrong.

Referring to \cref{fig:systemmodel}, the system model represents the beamer and sailcraft as two areas that are perpendicular to a common axis.  Transfer efficiency is a function of a single dimensionless parameter\footnote{Defined here as Goubau defines it \cite{goubau1961guided,goubau19683}. Others define it differently \cite{hansen2005universal,brown1992beamed}.} $\tau$ that depends on the product of optic diameters $D_{s}$ and $D_{b}$, the distance between them $z$, and the beam wavelength $\lambda$:
\begin{equation}
\tau \equiv 2\pi \frac{\lambda z}{\sqrt{\mathcal{A}_{s}\mathcal{A}_{b}}}=\frac{8\lambda z}{D_{s}D_{b}}.
\label{eq:goubautaudef}
\end{equation}
There is no closed-form solution for beam power transfer efficiency $\eta_b(\tau)$; however, it is closely approximated by \cite{parkin2017mtpfinalreport}:
\begin{equation}
\eta_b \left( a\right)=\left\lbrace 
\begin{array}{ll}
\eta _{1}\left( a\right) & \text{if }a>1.21748051194181 \\ 
\eta _{2}\left( a\right) & \text{otherwise}
\end{array} 
\right.
\label{eq:goubauefficiency}
\end{equation}
\begin{equation}
\eta _{1}\left( a\right) =\frac{1}{4b^2}\left( a^{4}+\allowbreak \sqrt{a^{8}-4a^{4}b+4b^{2}-8b+4}\allowbreak \right)^{2}
\end{equation}
\begin{equation}
\eta _{2}\left( a\right) =\left( \frac{a^{2}}{2}-\frac{a^{6}}{32}+\frac{7a^{10}}{4608}\right) ^{2},
\label{eq:goubaueta2}
\end{equation}
where
$a\left( \tau \right) \equiv \sqrt{\frac{2\pi }{\tau }}$
and
$b\equiv e^{a^{2}}$.

\begin{figure}[h]
\begin{tikzpicture}[ 
declare function={
	atau(\t)    = sqrt(2*pi/\t); 
	btau(\t)    = exp(2*pi/\t); 
	etao(\a,\b) = ((\a^4+sqrt(\a^8-4*\b*\a^4+4*\b^2-8*\b+4))/2/\b)^2; 
	etat(\a)    = (0.5*\a^2-(\a^6)/32+(\a^10)/4608*7)^2; 
	etas(\t)    = ((\t==0)||atau(\t)>1.2174805119418)*etao(atau(\t),btau(\t))+(\t!=0)*(atau(\t)<=1.2174805119418)*etat(atau(\t)); 
}]
\begin{axis}[
	axis x line=bottom,
	axis y line=left,
	grid=both,
    xlabel style={overlay},
	xlabel={$\tau$},
	ylabel={$\eta_b\left(\tau\right)$},
	domain=0:21,ymin=0,ymax=1.1,xmin=0,xmax=21,
	] %
\addplot[color=blue,samples=500]{etas(x)};
\end{axis}
\end{tikzpicture}
\caption{Goubau beam power transmission efficiency calculated using \cref{eq:goubauefficiency}}
\label{fig:goubau}
\end{figure}
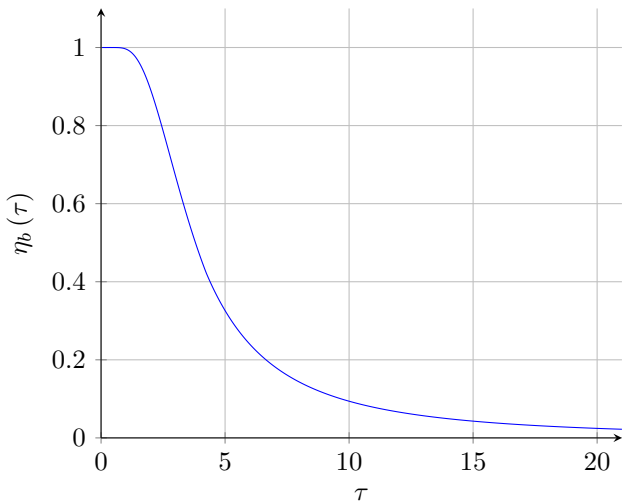
\Crefrange{eq:goubautaudef}{eq:goubaueta2} enable the rapid recalculation of power reaching the sail on each time-step of the trajectory integration via
\begin{equation}
P_{b}=\eta_a \eta_b P_{1},
\label{eq:goubauP0}
\end{equation}
where $\eta_a$ is the efficiency factor that accounts for atmospheric attenuation via absorption and scattering. For consistent accounting throughout this paper, $P_{1}$ is defined to be the laser power that is transmitted by the beamer. $P_{b}$ is therefore the fraction of transmitted power that is destined to reach the sailcraft.  It varies monotonically with $\tau$, a desirable feature when using numerical optimization.

\subsection{Equation of Motion}
The equation of motion relates the power that is incident upon the sailcraft to the sailcraft's acceleration.  It is deduced by requiring momentum to be conserved through the interaction of the sailcraft and beam photons. The derivation presented here extends the approach of Kulkarni \cite{kulkarni2016relativistic} to include a dielectric sailcraft having finite transmittance.  Also, the sailcraft thermally re-emits absorbed beam energy in the forward and backward directions, and this is included here because it contributes a nonzero drag force as seen from the beamer frame. 

In the beamer (observer) frame, conservation of momentum can be expressed as
\begin{equation}
\frac{dp_{p}}{dt}+\frac{dp_{s}}{dt}=0,
\label{eq:pconserved}
\end{equation}
where the relativistic sailcraft momentum $p_{s}$ is given by
\begin{equation}
p_{s}=\gamma m_{0}c\beta.
\end{equation}
$c$ is the speed of light, $m_{0}$ is the sailcraft rest mass, and $\gamma$ is the Lorentz factor $1/\sqrt{1-\beta ^{2}}$.  Newton's second law shows the time rate of change of sailcraft momentum to be equal the apparent force $F_{s}$ acting on the sailcraft as seen from the beamer frame,
\begin{equation}
F_{s}=\frac{dp_{s}}{dt}=\gamma ^{3}m_{0}c\dot{\beta}.
\label{eq:forceexpr}
\end{equation}
\begin{figure}[h]
\centering
\includegraphics[width=0.75\columnwidth]{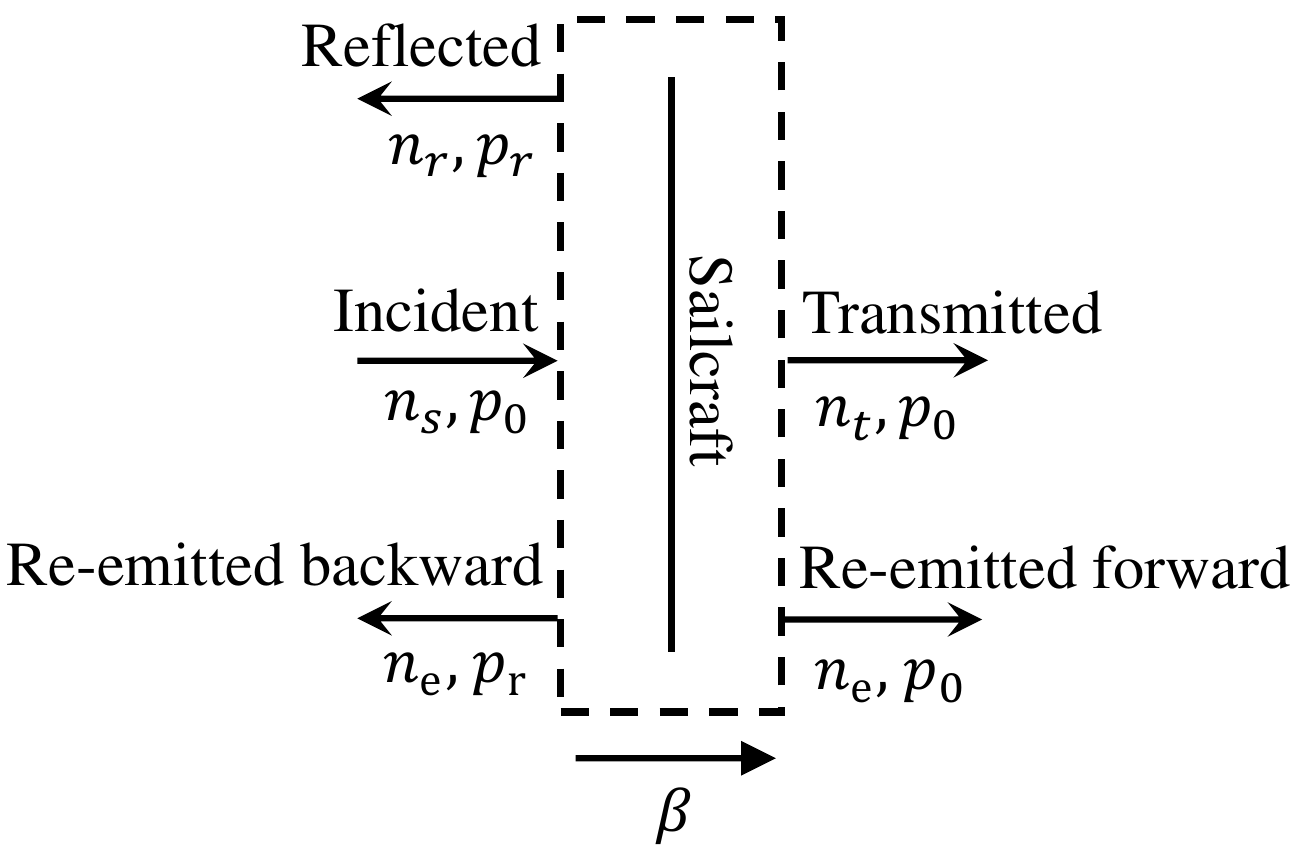}
\caption{Control volume}
\label{fig:controlvolume}
\end{figure}
The time rate of change of photon momentum is calculated using a control volume that co-moves with the sailcraft, shown in \cref{fig:controlvolume}. Photons cross the control boundary with rates $n$ and momenta $p$. In the figure, the arrows are drawn such that the rates and momenta are positive quantities. Of the photons which are destined to strike the sailcraft (not all or even most of them), let us say that $n_{b}$ is the rate at which they are transmitted by the beamer. The sailcraft recedes from the beamer at speed $\beta$.  Thus, the rate at which photons strike the sailcraft, $n_{s}$, is given by
\begin{equation}
n_{s}=n_{b}\left( 1-\beta \right).
\end{equation}
Consistent with the other rates, this rate is the number of photons per unit \textit{beamer time}, as opposed to sailcraft time. This is important because there is time dilation between the beamer and sailcraft frames. The rate $n_{r}$ at which photons are reflected is related to the reflectance $R$ by
\begin{equation}
n_{r}=Rn_{s}.
\label{eq:nreflectance}
\end{equation}
This model asserts that the sail is in thermal equilibrium because it re-emits the energy that it absorbs and has no thermal inertia. Thus, the rate $n_{t}$ at which photons are transmitted is given by
\begin{equation}
n_{t}=\left( 1-R-A\right) n_{s},
\end{equation}
where $A$ is the absorptance. Absorbed photons heat the sailcraft. To maintain thermal equilibrium, their energy is re-radiated in the forward and backward directions as photons spanning a range of wavelengths.  For accounting purposes, this model pretends that the incident photons that would have been absorbed are instead reflected and transmitted in equal measure, in a similar way to the other photons. Thus, the photons that are re-emitted are assigned momentum $p_{r}$ in the backward direction and $p_{0}$ in the forward direction. The rate $n_{e}$ at which photons are re-emitted is related to the absorptance by
\begin{equation}
n_{e}=\frac{A}{2}n_{s}.
\end{equation}

Photon momenta are unchanged by the moving control volume, but they are affected by frame conversions within.  Photons that are reflected from the sail are Doppler shifted twice:  First in going from the beamer to sailcraft frame, then after reflection in going back from the sailcraft to the beamer frame.  The photon is redshifted both times because the sailcraft is receding from the beamer and vice versa.  In the beamer frame, an incident photon has momentum $p_{0}$. This is related to the beamer wavelength $\lambda_{0}$ by $p_{0}=\frac{h}{\lambda_{0}}$, where $h$ is Planck's constant. Thus, the ratio of reflected to incident momentum is given by the inverse square of the relativistic Doppler factor:
\begin{equation}
\frac{p_{r}}{p_{0}}=-\frac{\lambda _{0}}{\lambda _{r}}=-\frac{1-\beta }{1+\beta }.
\label{eq:eqndopplershift}
\end{equation}

Summing the contributions at the boundary in \cref{fig:controlvolume},
\begin{equation}
\frac{dp_{s}}{dt}=n_{s}p_{0}-n_{t}p_{0}+n_{r}p_{r}+n_{e}p_{r}-n_{e}p_{0}.
\end{equation}
Finally, substituting \crefrange{eq:pconserved}{eq:eqndopplershift} into this expression yields the equation of motion:
\begin{equation}
F_{s}=\frac{P_{b}}{c}\frac{1-\beta }{1+\beta }\left( A+2R\right),
\label{eq:eqnmotion}
\end{equation}
where $P_{b}=c p_{0}n_{b}$ represents the beam power that is destined to strike the sailcraft as measured in the beamer frame.  At any point in the trajectory, $P_{b}$ is calculated from \cref{eq:goubauP0}. \Cref{eq:eqnmotion} can be simplified by recognizing that in the sailcraft rest frame, the sailcraft receives a power $P_{s}^{\prime }$ given by \cite{einstein1905elektrodynamik}
\begin{equation}
P_{s}^{\prime }=\frac{1-\beta }{1+\beta }P_{b}.
\label{eq:eqnthankyoueinstein}
\end{equation}
Thus,
\begin{equation}
F_{s}=\frac{P_{s}^{\prime }}{c}\left( A+2R\right).
\label{eq:eqnmotionsimple}
\end{equation}
Combining \cref{eq:eqnmotion} with \cref{eq:forceexpr} yields another form of the equation of motion that is more readily integrated,
\begin{equation}
\dot{\beta}=\frac{P_{b}}{\gamma ^{3}E_{0}}\frac{1-\beta }{1+\beta }\left( A+2R\right),
\label{eq:eqnmotionalternate}
\end{equation}
where $E_{0}=m_{0}c^{2}$ is the sailcraft rest energy.

If the derivation of force is repeated for only the re-emitted forward and backward photons shown in \cref{fig:controlvolume}, then a drag-like force is found:
\begin{equation}
F_{d}=-\frac{P_{a}^{\prime }}{c}\beta,
\label{eq:eqndrag}
\end{equation}
where the absorbed power $P'_{a}$ is given by
\begin{equation}
P'_{a}=A P'_{s}.
\label{eq:eqnpa}
\end{equation}
Physically, the thermally-radiated photons blueshift in the forward direction and redshift in the backward direction.  This unbalances photon momenta between the two directions and manifests as an apparent drag that acts on the sailcraft from the beamer frame, yet is not felt by the sailcraft in its rest frame.  In the context of dust particles that orbit stars, this force is known as Poynting-Robertson drag \cite{robertson1937dynamical}.  For a non-absorbing sailcraft, there is no drag, but for a perfectly absorbing sailcraft, the equations of Kulkarni \cite{kulkarni2016relativistic,kulkarni2018relativistic} overpredict the sailcraft's acceleration by a factor of $\left( 1+\beta \right) $, which is 20\% at \SI{0.2}{c}.
\begin{figure}[h]
\centering
\includegraphics[width=\columnwidth]{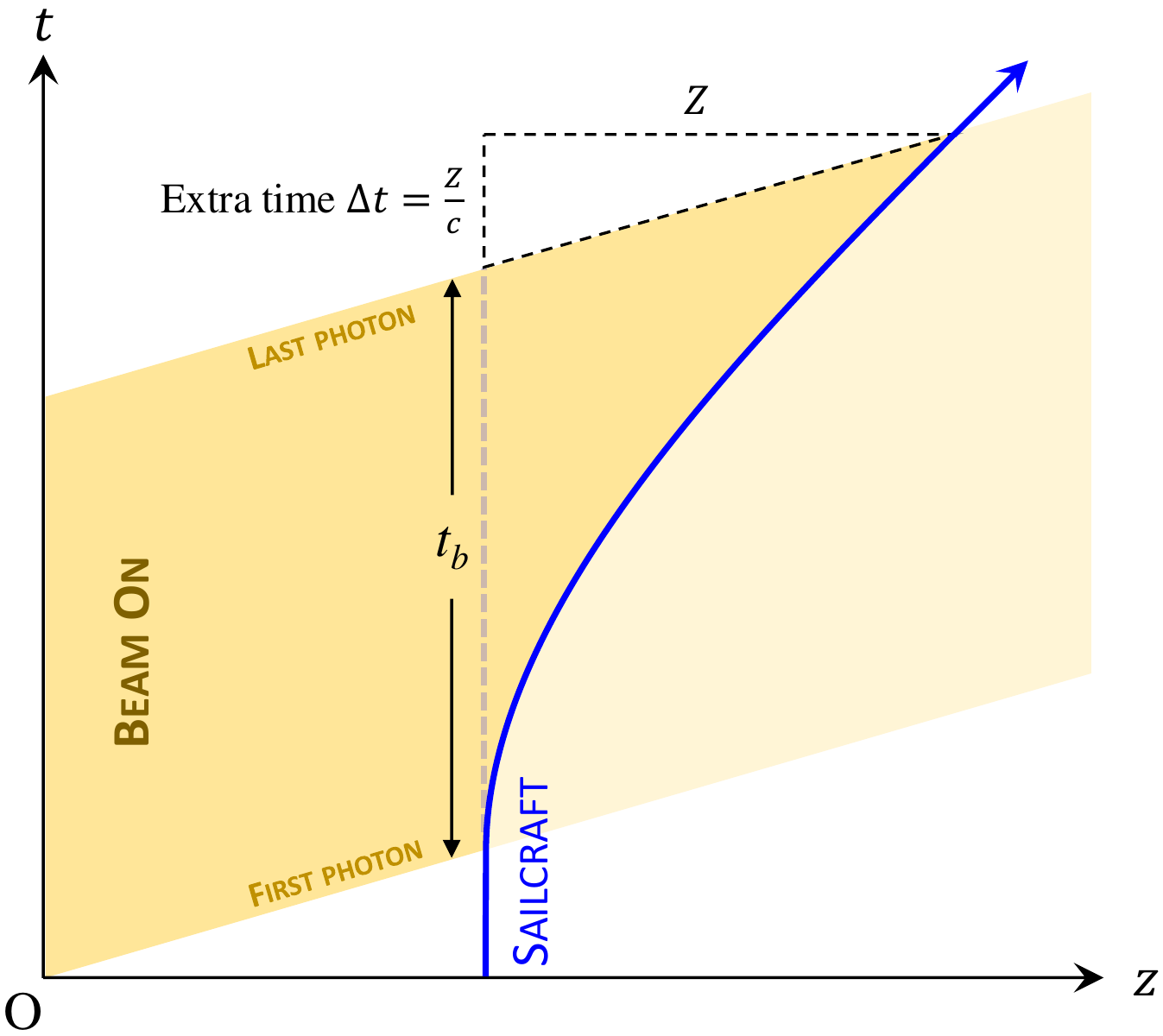}
\caption{Spacetime diagram}
\label{fig:spacetime}
\end{figure}

This derivation has said that photons strike the sailcraft at a rate that is slowed by the classical factor of $\left( 1-\beta \right)$, so where do the missing photons go? The situation is clarified by drawing a spacetime diagram, shown in \cref{fig:spacetime}. The missing photons simply catch up with the sailcraft after the beam is turned off.  Thus, if a sailcraft accelerates from rest until it reaches the desired cruise velocity at time $t_{s}$ and displacement $Z$, then the beam duration $t_{b}$ is given by
\begin{equation}
t_{b}=t_{s}-\frac{Z}{c}.
\label{eq:eqntbts}
\end{equation}
The sailcraft equilibrium temperature $T_{s}$ is estimated by using the Stefan-Boltzmann equation,
\begin{equation}
T_{s}=\root{4}\of{\frac{P'_{a}}{\sigma \varepsilon }},
\label{eq:eqnts}
\end{equation}
where $\sigma$ is the Stefan-Boltzmann constant and $\varepsilon$ is the 2-sided total hemispherical emittance.  This means that for a temperature-limited sailcraft, $P'_{s}$ is constant.  In the sailcraft rest frame, the equation of motion is simply
\begin{equation}
F_{s}^{\prime }=\frac{P_{s}^{\prime }}{c}\left( A+2R\right).
\label{eq:eqnmotionsailcraftframe} 
\end{equation}
Thus, $F_{s}^{\prime }$ is constant for a temperature-limited sailcraft.  

Equating \cref{eq:eqnmotionsailcraftframe} with \cref{eq:eqnmotionsimple} yields simply 
\begin{equation}
F_{s}^{\prime }=F_{s}.
\end{equation}
Thus, the force is the same in the sailcraft and beamer frames.  This is consistent with a pure force that does not add net heat to the sailcraft.  Other forms of the equation of motion produce impure forces that are frame-dependent and imply sailcraft heating.

The acceleration varies between sailcraft and beamer frames because of the factor of $\gamma ^{3}$ introduced by \cref{eq:forceexpr}. This variation is seen in the trajectory integration results, presented a little later.
\subsection{Stratified Layer Optical Model}
Which is better, a \SI{1000}{\nano\meter} thick sail with 99.9\% reflectance, or a \SI{100}{\nano\meter} thick sail of the same mass density, but with 25\% reflectance?  To define `better', a figure of merit is needed.  For simple design purposes, the appropriate figure of merit is not the sail reflectance or the sail thickness, but the sail acceleration per unit beam power.  Recalling the equation of motion, \cref{eq:eqnmotionalternate}, and simplifying with $A\ll R$ and $\beta \ll 1$ yields,
\begin{equation}
\frac{\dot{\beta}}{P_{b}}\propto \frac{R}{\rho \delta },
\end{equation}
because $\gamma \rightarrow 1$ and $E_{0}\propto \rho \delta $ where $\rho$ is sail mass density and $\delta$ is its thickness.  Thus, it can now be seen that the \SI{100}{\nano\meter} thick sail is better and accelerates 2.5 times faster than the alternative, all other things being equal.

The question of optimum sail thickness and performance is further complicated because dielectric layer reflectance varies with thickness.  The Starshot system model implements a complex characteristic matrix method as described by Macleod \cite{macleod2010thin} to calculate the reflectance, transmittance and absorptance of an arbitrary but locally-flat assembly of thin films.  

\begin{figure}[h]
\centering
\includegraphics[width=\columnwidth]{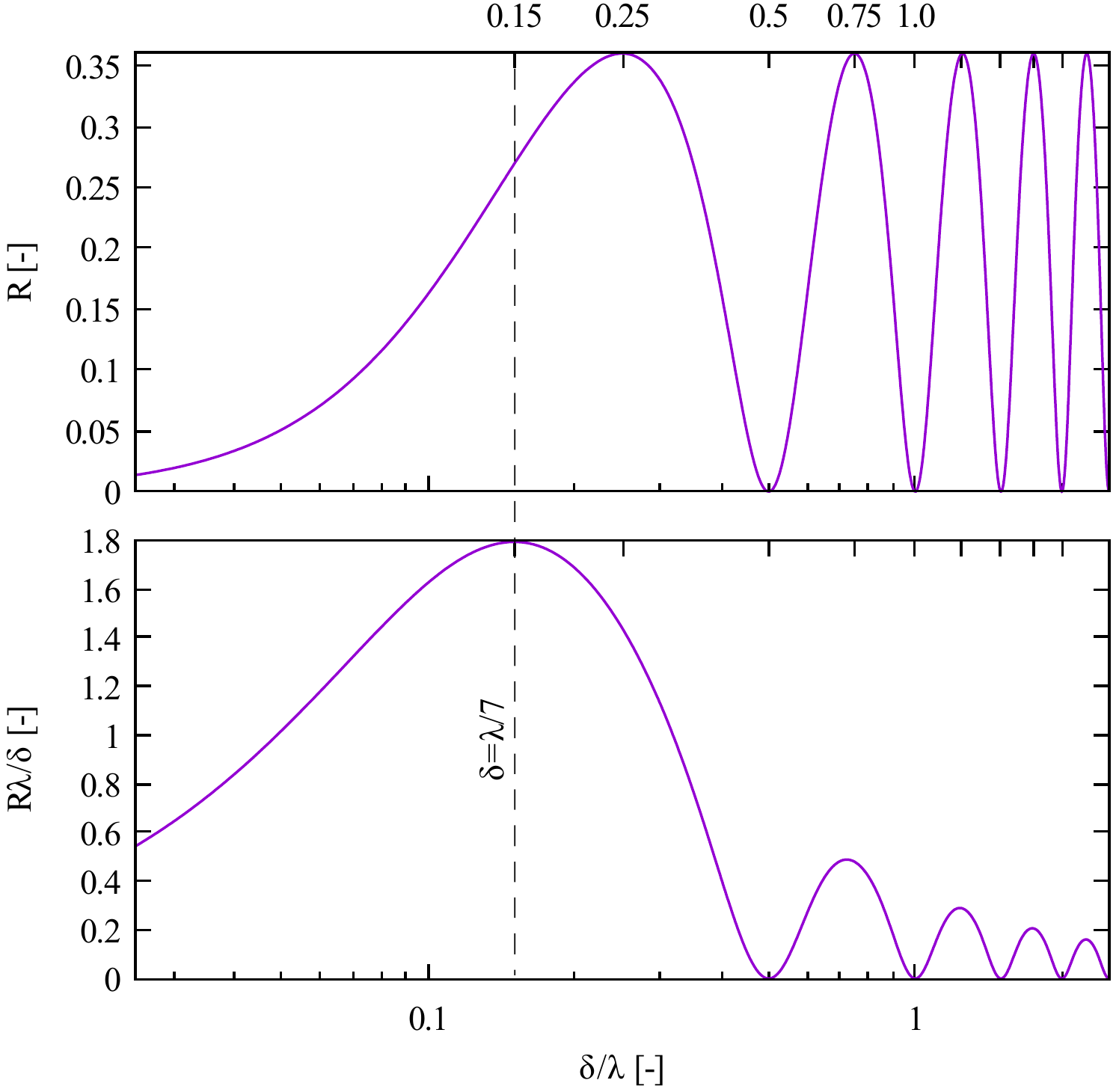}
\caption{Figures of merit predicted by the stratified layer model}
\label{fig:stratlayer}
\end{figure}

In \cref{fig:stratlayer}, the model is used to plot the reflectance of a single-layer dielectric film having a relative refractive index of 2 under illumination by free-space plane waves oriented at normal incidence to the film. Thus, the incident waves have half their free-space wavelength within the film, and the maximum reflectance occurs when the film is one quarter of this thickness, three quarters, five quarters, and so on. As already shown, system performance is maximized by choosing the thickness at which $R/\delta$ is maximum.  \Cref{eq:nreflectance} shows this to occur at about $\lambda/7$, with the subsequent maximum having more than three times worse performance.  For illumination at a free-space wavelength of \kpWavelength, this corresponds to an optimum sail thickness of \SI{76}{\nano\meter}. 

\section{Key Trade}

It is always possible to vary beam diameter, power, and duration such that a sailcraft reaches its desired cruise velocity.  Thus, the key trade lies in reaching the desired cruise velocity \textit{at minimum cost}. 

\begin{figure}[h]
\centering
\includegraphics[width=\columnwidth]{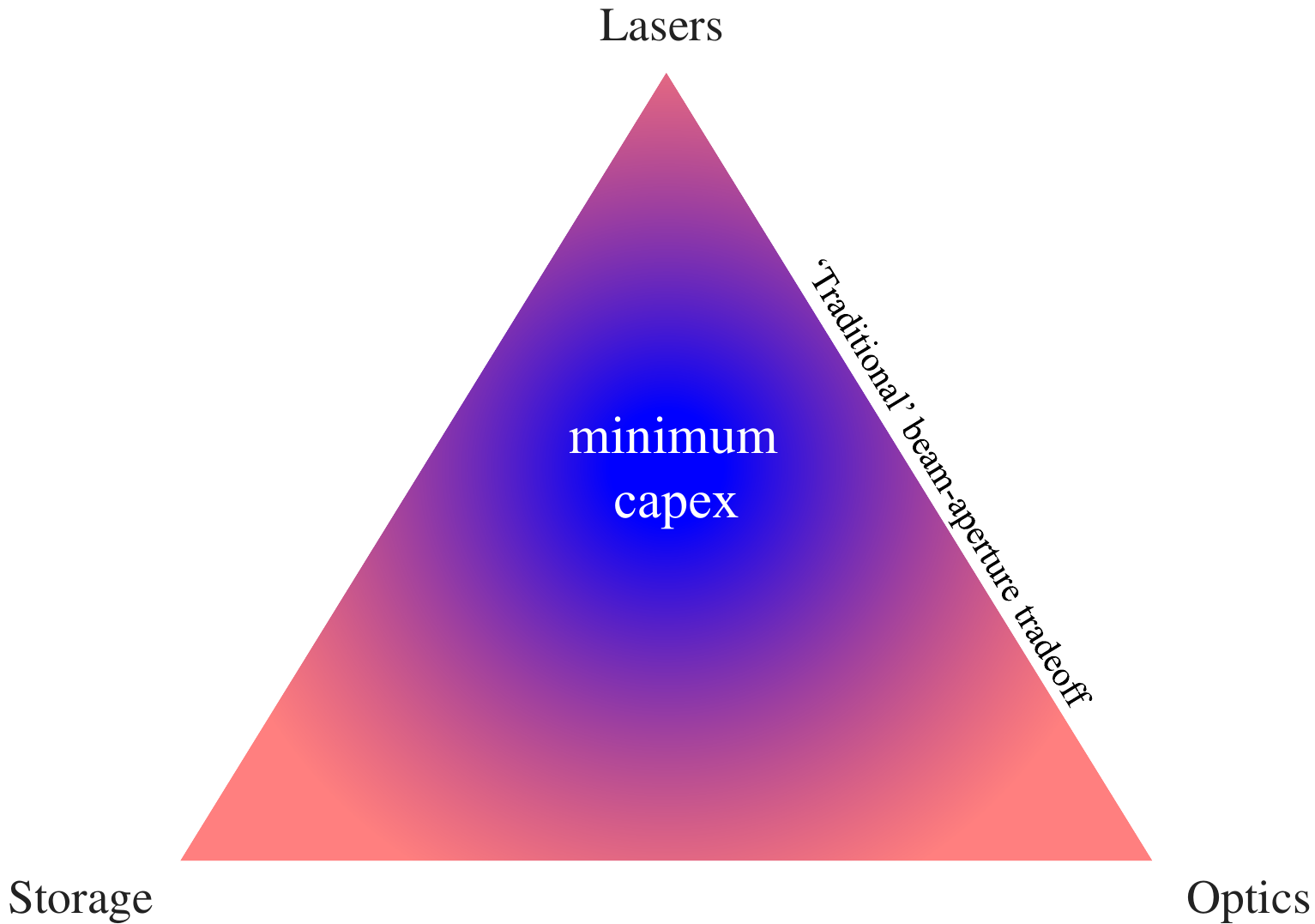}
\caption{Key trade}
\label{fig:keytrade}
\end{figure}

Benford \cite{benford2011costoptimizedsails} presents an analysis for cost-optimized beam driven sail missions in which the beamer's primary optic diameter is traded vs. power in order to minimize capex\footnote{capital expenditure}: A smaller diameter saves money on optics but reduces transfer efficiency, so power is increased to compensate.  Conversely, lower power saves money on sources but transfer efficiency has to be improved to compensate, by enlarging the primary optic. Somewhere between a beamer that is very powerful with a tiny primary and one that is very weak with a gigantic primary, there exists a happy medium that minimizes capex.  Benford's results affirm a rule of thumb used by microwave system designers for rough estimates: Minimum capex is achieved when the cost is equally divided between antenna gain and radiated power.  Similar results have been obtained in the field of beamed energy launch vehicles \cite{parkin2017mtpfinalreport,kare2006comparison}.  Hereafter, we refer to such analyses as ``traditional beam-aperture tradeoffs''.

The ``traditional beam-aperture tradeoff'' optimum sits on the axis between lasers and optics in \cref{fig:keytrade}. However, this is not a global minimum capex. This paper incorporates the key realization that beam duration is an independent variable that can be traded vs. power and diameter to lower the capex even further. Each of the three costs shown in \cref{fig:keytrade} can be traded for increases in the others:  Dominant laser cost is mitigated by reducing power in favor of longer beam duration (increases losses) and larger beamer diameter (decreases losses). Dominant optics cost is mitigated by reducing beamer diameter (increases loses) in favor of higher power.  Dominant energy storage cost is mitigated by improving the efficiency of energy transfer from the beamer to the sailcraft via increased beamer diameter and decreased beam duration.

A simple model for beamer capex $C$ is given by
\begin{equation}
C=k_{a}\frac{\pi D_{b}^{2}}{4}+k_{l}P_{1,max}+k_{e}Q_{0}.
\label{eq:eqncost}
\end{equation}
Beamer diameter $D_{b}$, peak transmitted power $P_{1,max}$, and pulse duration $t_{b}$ are dependent variables of the system model, as explained in the next section.  Stored energy $Q_{0}$ is the integral of the wallplug power drawn over the pulse duration, and this power draw is not constant, as shown later. Factors $k_{a}$, $k_{l}$, and $k_{e}$ are independent user-supplied values for cost per unit area, cost per unit power, and cost per unit energy stored. They are technology figures of merit.  By choosing a high laser cost factor of \SI[per-mode=symbol]{1000}[\$]{\per\watt}, for example, the cost-optimum solution moves toward longer beam duration, reduced average power, and larger diameter, to improve energy transfer efficiency at long range. At present, the cost model encompases the beamer only and not the sailcraft or other elements of the system. The beamer is expected to be the largest capital expense.

\section{Solution Procedure}
\begin{figure}[hbtp]
\centering
\includegraphics[width=\columnwidth]{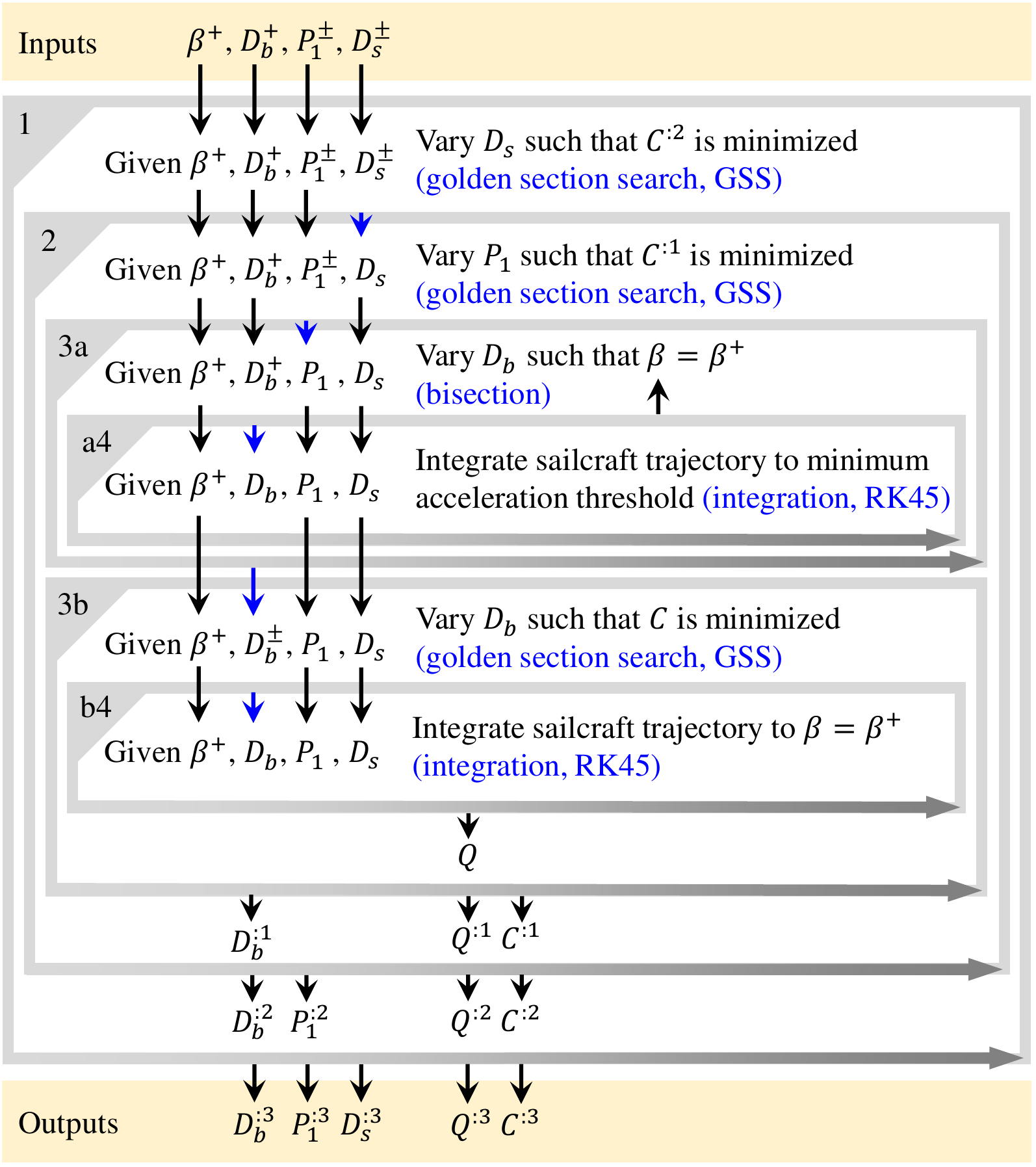}
\caption{Solution procedure}
\label{fig:solutionproc}
\end{figure}

A solution procedure for the system model is shown in \cref{fig:solutionproc}. At its top levels, the system model performs nested optimizations to ensure that the sailcraft reaches its cruise velocity and that it does so with system elements whose specifications minimize the beamer capex.  

Starting at the top of \cref{fig:solutionproc}, the desired cruise velocity $\beta ^{+}$ is specified along with upper and lower bounds for the beamer power, $P_{1}^{\pm }$, and sailcraft diameter $D_{s}^{\pm }$.  Only a maximum beamer diameter $D_{b}^{+}$ is specified.  These values are brought into the outermost iteration (iteration 1), a golden section search varying $D_{s}$ between its chosen bounds $D_{s}^{\pm }$ such that beamer capex $C^{:2}$ is minimized.  The `:2' in the superscript means that $C$ is already twice optimized (with respect to $P_{1}$ and $D_{b}$).  Shown at the bottom of the figure, iteration 1 returns the cost-optimal sailcraft diameter $D_{s}^{:3}$ together with the corresponding capex $C^{:3}$, now optimal over three dimensions. Before these optimal values are returned, all the internal optimizations must first run, and these draw on the optimizer-chosen value of $D_{s}$, which is passed inwards to iteration 2.

Iteration 2 is a golden section search varying $P_{1}$ between its chosen bounds $P_{1}^{\pm }$ such that capex $C^{:1}$ is minimized.  The `:1' in the superscript means that it is already optimized with respect to $D_{b}$ only. This iteration returns $P_{1}^{:2}$.  It also returns the corresponding capex $C^{:2}$ to iteration 1, which uses its value to guide the optimizer in its choice of $D_{s}$ for the next iteration. The newly-defined value of $P_{1}$ is passed inwards to iterations 3a and 3b.
 
Iteration 3a is a bisection solver varying $D_{b}$ between zero and its upper bound of $D_{b}^{+}$ such that sailcraft velocity at the end of trajectory integration, $\beta$, is only just equal to its desired value of $\beta ^{+}$.  By setting $D_{b}^{-}$ in this way, the solution procedure assures that beam cutoff always occurs as the sailcraft reaches exactly its desired velocity $\beta ^{+}$.  It remains to minimize the capex.

Iteration a4 is a trajectory integration (implemented using the RK45 algorithm) returning the sailcraft velocity $\beta$ at the point that its acceleration falls below a minimum threshold, which terminates integration. 

Iteration 3b is a golden section search varying $D_{b}$ between its chosen upper bound $D_{b}^{+}$ and its calculated lower bound $D_{b}^{-}$ such that capex $C$ is minimized.  As described by cost \cref{eq:eqncost}, $C$ is an implicit function of the sailcraft trajectory, so this trajectory must be recalculated on each iteration. Iteration 3b returns $D_{b}^{:1}$ and its corresponding capex $C^{:1}$ to iteration 2, which uses $C^{:1}$ to guide the optimizer in its choice of $P_{1}$ for the next iteration.

Iteration b4 is a trajectory integration (implemented using the RK45 algorithm) that returns the pulse energy $Q$ because it is needed in cost \cref{eq:eqncost} to compute energy storage cost.

Finally, the result of all these iterations are values for three dimensions; $D_{b}^{:3}$, $P_{1}^{:3}$, and $D_{s}^{:3}$; that produce a minimal capex $C^{:3}$ and cost-optimal values for $Q^{:3}$.  All these values are used in conjunction with auxiliary equations to quantify point designs.      

\paragraph{Simplified method for constant payload fraction} If payload were a constant fraction of sail mass, the solution procedure could be simplified by omitting the outer iteration on $D_{s}$.  Beamer intensity $I_{b}$ would then be varied such that $C/\mathcal{A}_{b}$ is minimized (instead of $P_{1}$ varied such than $C$ is minimized).  Also, $\tau$ would be optimized instead of $D_{b}$.  

Given cost-optimal values for $C/\mathcal{A}_{b}$, $I_{1}$, and $\int I_{1}dt$, a separate optimization could then vary $D_{s}$ to minimize $C$.  This optimization would iterate the Goubau and cost models without recalculating the computationally-expensive trajectory. This amounts to a `traditional' power vs. aperture tradeoff as shown in \cref{fig:keytrade}.  Thus, it is inferred that storage is an axis of the tradespace only because payload mass is not a constant fraction of sail mass.  Also, constant costs that do not scale with beamer area $C/\mathcal{A}_{b}$ in cost \cref{eq:eqncost} would again unfold storage as an axis of the tradespace.
\section{Point Designs}
\subsection{\SI{0.2}{c} Mission}
The \SI{0.2}{c} point design embodies key elements of mid-21st century Starshot missions to nearby stars, including those to the Centauri System. 
\subsubsection{Inputs}
The inputs to the mission point design are summarized in \cref{tab:02cinputs}. A \kpWavelength\ wavelength is consistent with ytterbium-doped fiber amplifiers.  An initial sailcraft displacement of \SI{60000}{\kilo\meter} is consistent with a low-thrust non-Keplarian orbit \cite{forward1984light,baig2010light} that keeps the sailcraft (and associated spacecraft) stationary in the sky relative to the target star.  
\begin{table}[h]
\caption{System model inputs for \SI{0.2}{c} mission}
\label{tab:02cinputs}
\centering
\begin{tabular}{l}
\hline			
\SI{0.2}{c} target speed \\
\kpWavelength\ wavelength \\
\SI{60000}{\kilo\meter} initial sail displacement from laser source \\
\\
\SI{1}{\gram} payload \\
\SI{0.2}{\gram\per\meter\squared} areal density \\
\kpAbsorptance\ spectral normal absorptance at \kpWavelength\\
70\% spectral normal reflectance at \kpWavelength\\
\kpMaxTemp\ maximum temperature\\
\kpEmittance\ total hemispherical emittance (2-sided, \kpMaxTemp)\\
\\
\SI[per-mode=symbol]{0.01}[\$]{\per\watt} laser cost \\
\SI[per-mode=symbol]{500}[\$]{\per\meter\squared} optics cost\\
\SI[per-mode=symbol]{50}[\$]{\per\KWH} storage cost\\ 
50\% wallplug to laser efficiency\\
70\% of beam power emerging from top of atmosphere\\
\hline  
\end{tabular}
\end{table}

A \SI{1}{\gram} payload is bookkept separately from sail mass and reserved for scientific instrumentation and associated support systems.  Sail mass is calculated by the system model using the value of $D_{s}$ chosen by the optimizer combined with the input areal density.  If the sail were an optimal-thickness layer of silicon dioxide, for example, it would have an areal density closer to \SI{0.3}{\gram\per\meter\squared}, a room temperature absorptance of less than $10^{-10}$ \cite{wandel2005attenuation}, and a normal reflectance of 12\%.  This point design uses a lower areal density of \SI{0.2}{\gram\per\meter\squared}, higher absorptance of \kpAbsorptance, and a higher reflectance of 70\%, consistent with a hot two-dimensional nanohole photonic crystal that is somewhat tuned for high reflectance \cite{atwater2018materials}.  Because the system model does not yet incorporate a photonic crystal model, the stratified layer optical model is turned off and absorptance and reflectance remain constant throughout trajectory integration.

A maximum temperature of \kpMaxTemp\ is placed on the sail to prevent thermal runaway and to prevent damage to non-operating electronic or photonic components that are part of the sail or its payload.  An assumed total hemispherical emittance of \kpEmittance\ accounts for the energy that is radiated by the sail in both the forward and backward directions at \kpMaxTemp. This emittance is much lower than that of typical materials because the sailcraft is so thin and because its absorptance is nearly zero at \kpWavelength.  In comparison, a blackbody at \kpMaxTemp\ emits most strongly at \SI{4.6}{\micro\meter}.  This means that to approach the radiative performance of a blackbody, the sail must switch from virtually invisible to strongly absorbing/emitting within as small a wavelength range as possible.

Cost factors are chosen such that beamer capex is less than \$10B.  In particular, the laser cost is \SI[per-mode=symbol]{0.01}[\$]{\per\watt}, four orders of magnitude lower than the present cost, consistent with nearly automated production and high yields.  Automated production lines for microwave oven magnetrons long ago reached this cost.  The optics cost is \SI[per-mode=symbol]{500}[\$]{\per\meter\squared}, three orders of magnitude lower than the present cost for diffraction-limited optics, again consistent with nearly automated production and high yields.  This value is comparable with the retail cost of computer monitors.  \SI[per-mode=symbol]{50}[\$]{\per\KWH} energy storage is greater than current materials costs for some, but not all, energy storage technologies, and is more optimistic than current experience curve projections of future electrical energy storage costs \cite{schmidt2017future}.
\subsubsection{Results}
Upon running the system model using the inputs in \cref{tab:02cinputs}, the optimizers converge to the values given in \cref{tab:02coutputs}. The sail diameter that minimizes beamer capex is found to be \kpZeroTwoCSailDiameter, corresponding to a sailcraft mass of \kpZeroTwoCSailMass. At \SI{0.2}{c}, this mass has a relativistic kinetic energy of \kpZeroTwoCRelKE, whereas the beamer expends \kpZeroTwoCStored, corresponding to \kpZeroTwoCSysEfficiency\ system energy efficiency.  This energy costs only \kpZeroTwoCEnergy\ at a price of \SI[per-mode=symbol]{0.1}[\$]{\per\KWH}, making the energy cost three orders of magnitude lower than the \kpZeroTwoCCapex\ beamer capex.

\begin{table}[hbtp]
\caption{System model outputs for \SI{0.2}{c} mission}
\label{tab:02coutputs}
\centering
\begin{tabular}{l}
\hline			
\kpZeroTwoCCapex\ beamer capex comprised of: \\
\qquad\kpZeroTwoCLasers\ lasers (\kpZeroTwoCPowerMax\ max. transmitted power)\\
\qquad\kpZeroTwoCOptics\ optics (\kpZeroTwoCBeamerDiameter\ primary effective diameter)\\
\qquad\kpZeroTwoCStorage\ storage (\kpZeroTwoCStored\ stored energy)\\
\\
\kpZeroTwoCEnergy\ energy cost per Starshot (\kpZeroTwoCStored\ @\SI[per-mode=symbol]{0.1}[\$]{\per\KWH})\\
\kpZeroTwoCSysEfficiency\ system energy efficiency\\
\\
\kpZeroTwoCSailDiameter\ sail diameter\\
\kpZeroTwoCSailMass\ sailcraft mass (includes payload mass)\\
\\
\kpZeroTwoCDurationPulseMin\ (\kpZeroTwoCDurationPulseSec) beam transmit duration\\
\kpZeroTwoCDurationAccnMin\ (\kpZeroTwoCDurationAccnSec) sailcraft acceleration duration\\
\\
\kpZeroTwoCPhotonPressure\ temperature-limited photon pressure\\
\kpZeroTwoCPhotonForce\ temperature-limited force\\
\kpZeroTwoCAccnInit\ temperature-limited acceleration\\
\kpZeroTwoCAccnFin\ final acceleration (\kpZeroTwoCDispFinAu), \kpZeroTwoCDispFinLs\ from source\\
\\
\kpZeroTwoCFluxBeamerMax\ beamer maximum beam radiant exitance\\
\kpZeroTwoCFluxSailTLimited\ sailcraft temperature-limited irradiance\\

\hline	
\end{tabular}
\end{table}

\begin{figure*}[t]
\centering
\includegraphics[width=0.9\textwidth,keepaspectratio=true]{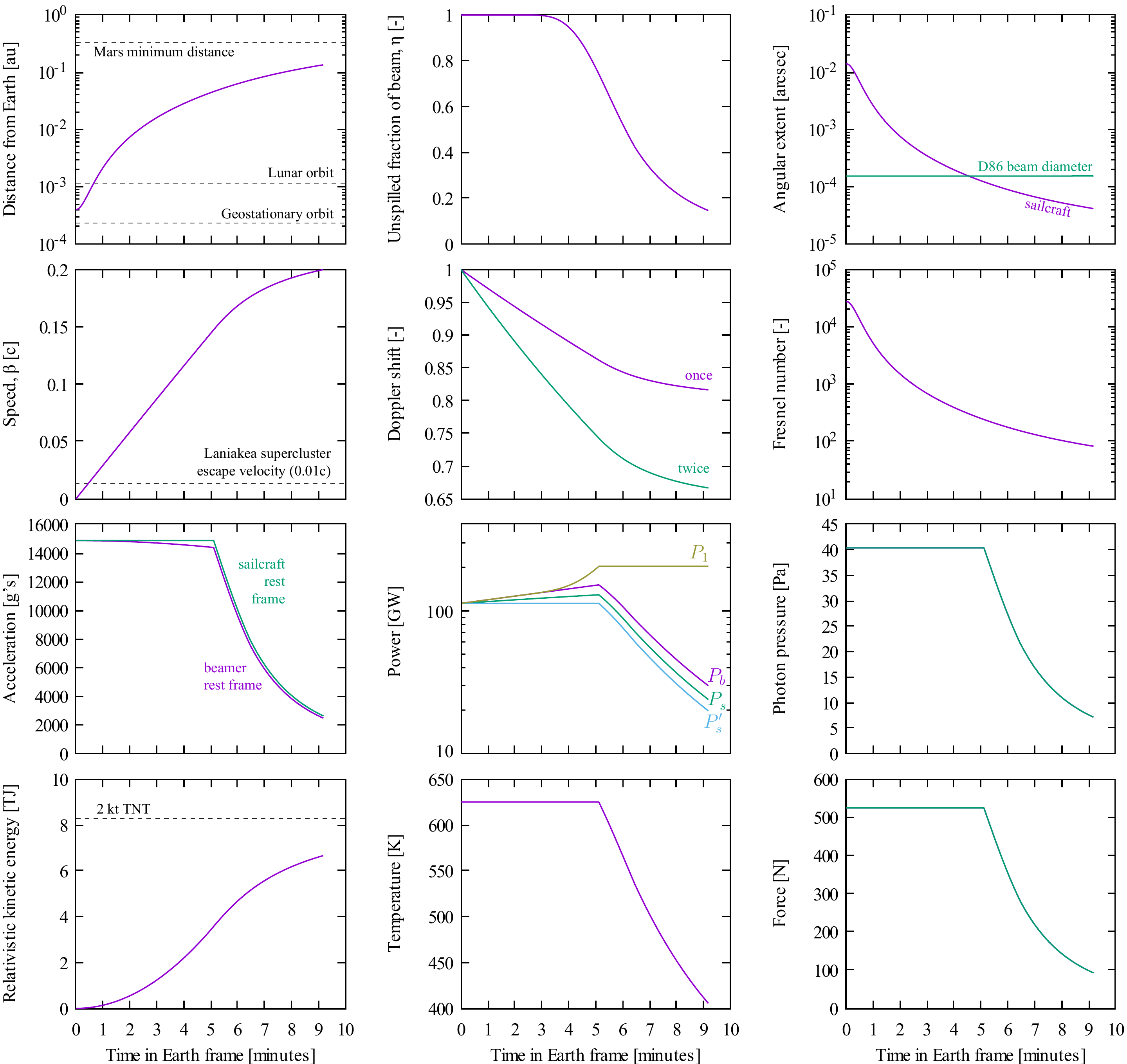}
\caption{\SI{0.2}{c} trajectory and related quantities}
\label{fig:02ctrajectory}
\end{figure*}

The \kpZeroTwoCCapex\ beamer capex is comprised of three unequal expenses with, surprisingly, laser cost being the smallest at \kpZeroTwoCLasers.  This is for a \kpZeroTwoCPowerMax\ maximum transmitted power, which is far below the petawatt-class lasers now in existence but many orders of magnitude greater in pulse energy (and duration).  The stored pulse energy of \kpZeroTwoCStored\ contributes the most to the beamer capex at \kpZeroTwoCStorage, but is only one fortieth of the \SI{2.5}{TWh} annual market for electrical energy storage that is projected for 2040 \cite{schmidt2017future}, with electric vehicles being the dominant source of demand.  Though it would be impractical to convene two million electric vehicle owners and drain their batteries for each starshot, it may be possible to use second-hand or donated battery packs that no longer perform well enough for electric vehicles.  

The remainder of the beamer capex is incurred by the optics; a filled array of telescope elements that form a \kpZeroTwoCBeamerDiameter\ effective diameter.  This effective diameter is the same as that of the Crescent Dunes solar energy concentrator, located in the California desert. Of course, the array modules for the beamer will be very different from solar concentrator mirrors.  When operating at maximum power, the beamer's radiant exitance is \kpZeroTwoCFluxBeamerMax\ (a spatial average obtained by diving the power output by the effective area of the primary optic). This exitance is forty times that of the solar concentrator operating at its peak, but three orders of magnitude lower than that of military laser beam directors.  In the plane of the sail, the beam converges to a much higher irradiance.  By limiting the beamer's power output, the system model reduces the irradiance at the sail to its temperature-limited value of \kpZeroTwoCFluxSailTLimited, which is three orders of magnitude lower than the flux of a \SI{1}{\kilo\watt} laser in a \SI{10}{\micro\meter} fiber, and five orders of magnitude lower than the non-thermal ablation threshold \cite{stuart1995laser}.

The \kpZeroTwoCFluxSailTLimited\ sail irradiance produces only \kpZeroTwoCPhotonPressure\ photon pressure, equivalent to a moderate breeze. But the sail is very thin and the breeze moves at the speed of light, resulting in \kpZeroTwoCAccnInit\ initial acceleration.  Such acceleration is experienced by bullets and artillery shells, but for fractions of a second. Even as the sailcraft reaches \SI{0.2}{c} after \kpZeroTwoCDurationAccnMin, it is still accelerating at \kpZeroTwoCAccnFin.  Such is the cost-optimum truncation point for the trajectory.

During the sailcraft's acceleration, a satellite that transits the beam in low Earth orbit would see a flash of little more than \kpZeroTwoCFluxBeamerMax\ for a fraction of second. This is because the high-flux part of the beam is the focus, and the focus is initially past geostationary orbit and only gets further away as the sail moves off. The beam and the ultra-high-acceleration sailcraft, behaving as if a speck of dust in optical tweezers, may be able to dodge satellites in medium Earth orbits or supersynchronous orbits. As is current practice, beam-satellite conjunction analyses would be performed to help schedule times at which lasing is allowable.  Also, there would be interlocks to dim or douse the beam if needed to protect unexpected aircraft or flocks of birds.  If the interruption were short enough, the trajectory could resume.

\begin{figure*}[t]
\centering
\includegraphics[width=0.9\textwidth]{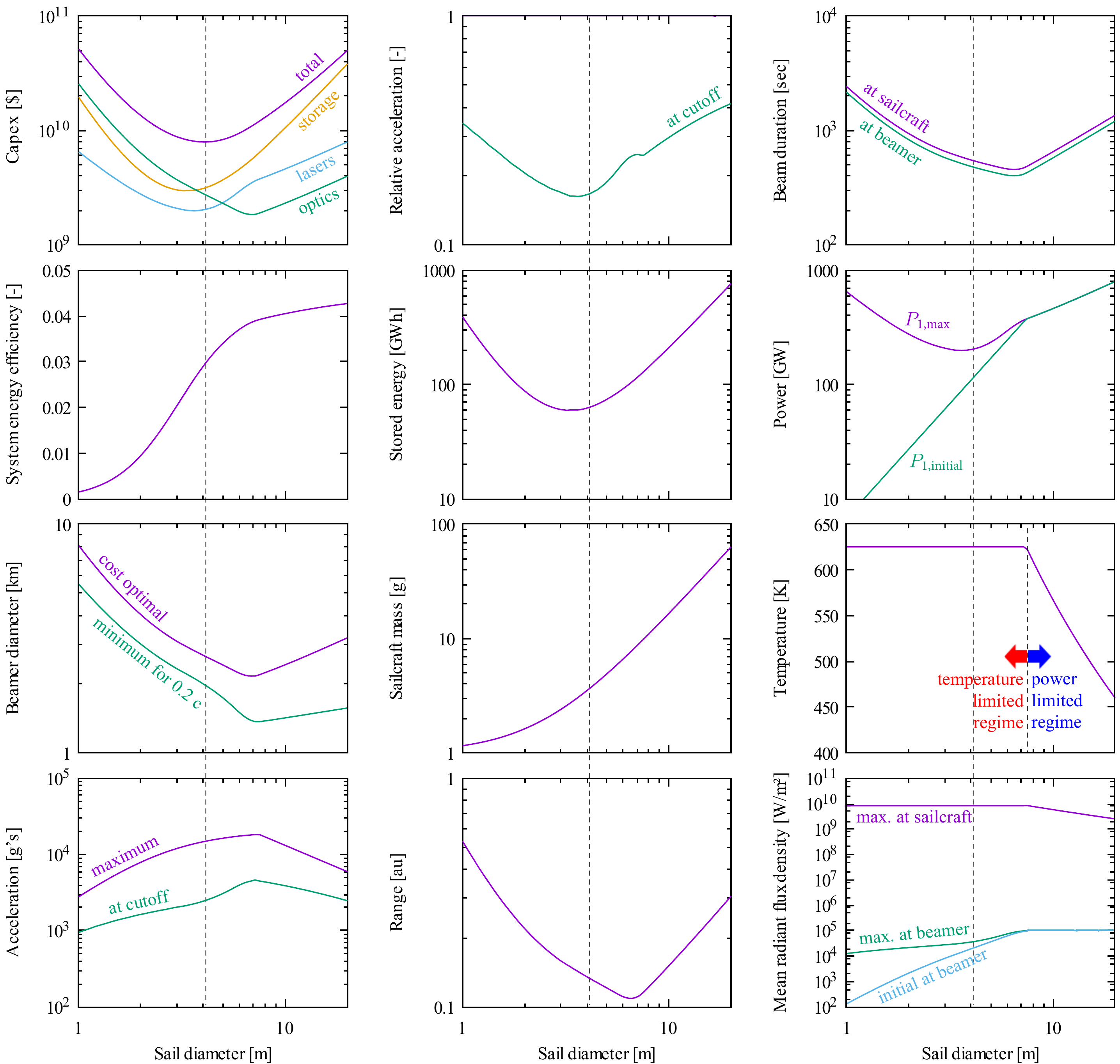}
\caption{\SI{0.2}{c} mission variation with sail diameter}
\label{fig:02cmissionsaildia}
\end{figure*}

The 1-D sailcraft trajectory and associated quantities are plotted as functions of time in \cref{fig:02ctrajectory}.  This trajectory corresponds to the cost-optimal point design summarized in \cref{tab:02coutputs}.  Starting at the top left plot, the sailcraft begins accelerating from a distance that would be just past the radius of geostationary orbit.  If the destination star were in the plane of the solar system, then the sailcraft would pass the Moon's orbit within the first minute, and could be a third of the way to Mars\footnote{at its minimum distance from Earth} by the end of acceleration around the \kpZeroTwoCDurationAccnMin\ mark.  Alpha Centauri is at \SI{-60}{\degree} declination, so its trajectory would be downwards out of the plane of the solar system.

The sailcraft speed increases almost linearly for the first half of the acceleration time.  Thereafter, the speed rises more gradually, until cutoff occurs precisely as \SI{0.2}{c} is reached. 

Essentially none of the beam is spilled until the trajectory reaches the \SI{3}{\minute} mark.  Until this time, the beamer power is throttled back to prevent the sail from overheating.  In the sail frame, this means that the incident and absorbed power, force, photon pressure and acceleration are all constant.  As seen from the beamer frame, relativistic effects cause the apparent acceleration to pull slightly downward as speed increases.

After the \SI{3}{\minute} mark, the beam begins to spill around the sail as it moves further away.  The beamer power gradually ramps up to compensate for increasing losses.  Evidently, it minimizes beamer capex to oversize the beamer power.  At some point, the beamer can no longer compensate for the increasing beam spillage, and the sail cools. By the end of the trajectory, less than 20\% of the transmitted power reaches the sail.  

Note that $P_{1}$ as plotted in \cref{fig:02ctrajectory} does not include finite photon travel time between the beamer and sailcraft, hence its time index is distorted.  The undistorted plot would be shifted and compressed left, then drop to zero at \kpZeroTwoCDurationPulseMin.  To avoid the need to post-calculate the undistorted $P_{1}$, pulse energy $Q_{0}$ is obtained by integrating $\left( 1-\beta \right) P_{1}$ along with the trajectory.

The sail's angular extent as seen from the beamer is initially within the pointing stability of telescopes such as the Hubble Space Telescope, but exceeds this as the sail gets further away. Even at the speed of light, the round trip time between the beamer and sail varies from \SI{0.4}{\second} at the beginning of the trajectory to \SI{134}{\second} at the end.  Given also the high sail accelerations, it is obvious that the beamer cannot actively point to follow the sail.  Instead, the sailcraft must be beam riding, seeking the axis of the beam by active or preferably passive stabilization schemes.

The Doppler shift plot shows the \kpWavelength\ beam redshifting as seen from the sailcraft frame, eventually reaching \SI{1.30}{\micro\meter} wavelength. The reflected light again redshifts, eventually reaching \SI{1.60}{\micro\meter} wavelength.  If the sailcraft were to transmit at the \SI{0.85}{\micro\meter} wavelength commonly used by VCSEL laser diodes, it would be received on Earth at \SI{1.04}{\micro\meter}.

The sailcraft's relativistic kinetic energy reaches \kpZeroTwoCRelKETJ\ (\kpZeroTwoCRelKE).  Per unit mass, this is \SI{2}{\peta\joule\per\kilo\gram}. In comparison, the heat produced by Pu-238 alpha decay is three orders of magnitude lower at \SI{2}{\tera\joule\per\kilo\gram}.  One way to tap the sailcraft's kinetic energy could be via the interstellar medium: The Local Interstellar Cloud is primarily composed of \SI{0.3}{atoms\per\centi\meter\cubed} of partially ionized hydrogen \cite{frisch2011interstellar}.  From the sailcraft's perspective traveling at cruise velocity, this manifests as a monochromatic hydrogen beam that is incident from the direction of travel, having a combined kinetic energy of \SI{55}{\watt\per\meter\squared}, or \SI{0.7}{\kilo\watt} over the sail's area if it faces the direction of travel.

In the highly unlikely event that a sailcraft were to collide with a planetary atmosphere, then the energy released would be equivalent to nearly \SI{2}{kt} of TNT. On average, one asteroid per year enters the Earth's atmosphere with this energy, though asteroids are orders of magnitude slower and heavier.  The sailcraft would vaporize before it got nearly as low into the atmosphere as asteroids do.

\subsubsection{Variation with respect to sail diameter}

Sail diameter is a variable that the system model varies to minimize beamer capex.  As shown in \cref{fig:solutionproc}, it is varied by the GSS algorithm of the outermost iteration (iteration 1).  By turning off this iteration, the design space as seen by the optimizer is plotted in \cref{fig:02cmissionsaildia}. Starting at the top left plot, the beamer (total) capex is minimum at the expected sail diameter of \kpZeroTwoCSailDiameter.  On every plot, this diameter is marked by a vertical line.  The line shows that the cost optimum does not exactly correspond to any other extrema; the optimum is a true tradeoff between storage, lasers and optics, as depicted in \cref{fig:keytrade}.

A striking feature in most of the plots in \cref{fig:02cmissionsaildia} is that the mission parameters vary in a qualitatively different way as sailcraft exceed \SI{8}{\meter} diameter.  This is because smaller sailcraft operate in a temperature-limited regime, whereas larger sailcraft operate in a beamer flux-limited regime.  On each trajectory integrator timestep, the system model calculates the sail temperature resulting from maximum beamer power.  If this temperature exceeds the maximum, this maximum becomes the boundary condition, and the corresponding beamer power (less than its maximum) is calculated.

Intuitively, one might expect that, relative to the baseline point design in \cref{tab:02coutputs}, money is saved by using a larger sail that is slowly accelerated by low power coming from a larger beamer over a longer period of time.  This limit corresponds to the right-hand side of each plot in \cref{fig:02cmissionsaildia}.  For sails that exceed \SI{8}{\meter} diameter, the optimum beamer diameter does indeed increase with sail diameter, as does beam duration and range.  Consistent with expectations, acceleration decreases in this limit. However, the laser power that is needed increases instead of decreases in the large sail diameter limit, and so capex also increases due to the laser cost and the increased cost of energy storage.  Thus, economics does not favor the slow acceleration school of thought.

\subsubsection{Variation with respect to technology figures of merit}

The \SI{0.2}{c} point design assumes particular cost performance figures of merit for the laser, optics and storage.  These are referred to as $k_{l}$, $k_{a}$, and $k_{e}$ in the simple cost model of \cref{eq:eqncost}.  But what happens if the lasers' cost performance, for example, is lower than expected?  Also, what happens if the material absorbs more, or less than assumed, or the payload is heavier or lighter?  

\begin{table}[h]
\caption{Relative impact of technology over/under performance}     
\label{tab:02ctechvariations}
    \begin{tabular}{ l | l | l | l | l }
     & Capex & $D_{b}$ & $P_{1}$ & $D_{s}$\\ \hline   
    \SI[per-mode=symbol]{0.1}[\$]{\per\watt} laser (10x) & 2.4 & 1.5 & 0.46 & 0.82\\    
    \SI[per-mode=symbol]{5000}[\$]{\per\meter\squared} optics (10x) & 2.4 & 0.52 & 2.5 & 1.4\\    
    \SI[per-mode=symbol]{500}[\$]{\per\KWH} storage (10x) & 3.5 & 1.6 & 0.83 & 0.77\\
    $10^{-7}$ absorptance (10x) & 4.1 & 2.3 & 1.5 & 1.8\\
    \SI{10}{\gram} payload (10x) & 2.7 & 1.3 & 2.3 & 1.7\\
    \hline
    \SI[per-mode=symbol]{0.001}[\$]{\per\watt} laser (0.1x) & 0.70 & 0.82 & 2.0 & 1.0\\
    \SI[per-mode=symbol]{50}[\$]{\per\meter\squared} optics (0.1x) & 0.48 & 2.0 & 0.45 & 0.74\\
    \SI[per-mode=symbol]{5}[\$]{\per\KWH} storage (0.1x) & 0.56 & 0.80 & 1.0 & 1.1\\
    $10^{-9}$ absorptance (0.1x) & 0.62 & 0.67 & 1.2 & 0.45\\
    \SI{0.1}{\gram} payload (0.1x) & 0.63 & 0.89 & 0.68 & 0.58\\
    \hline
    0.25 reflectance (0.36x) & 3.3 & 1.9 & 2.1 & 1.2\\
    0.90 reflectance (1.3x) & 0.76 & 0.87 & 0.84 & 0.96\\
    0.99 reflectance (1.4x) & 0.69 & 0.82 & 0.79 & 0.94\\
    \hline
    \SI{0.1}{c} cruise speed (0.5x) & 0.29 & 0.49 & 0.35 & 0.73\\
    \SI{0.4}{c} cruise speed (2x) & 4.7 & 2.3 & 4.0 & 1.3\\
    \SI{0.9}{c} cruise speed (4.5x) & 14 & 13 & 120 & 2.2\\
    \SI{0.99}{c} cruise speed (5x) & 1400 & 44 & 1600 & 2.7\\
    \hline
    $10^{-6}$ absorptance ($10^{2}$x) & 30 & 6.4 & 2.2 & 3.3\\
    $10^{-5}$ absorptance ($10^{3}$x) & 280 & 20 & 2.5 & 6.0\\
    \hline
    \end{tabular}
\end{table}

\begin{figure*}[t]
\centering
\includegraphics[width=0.9\textwidth]{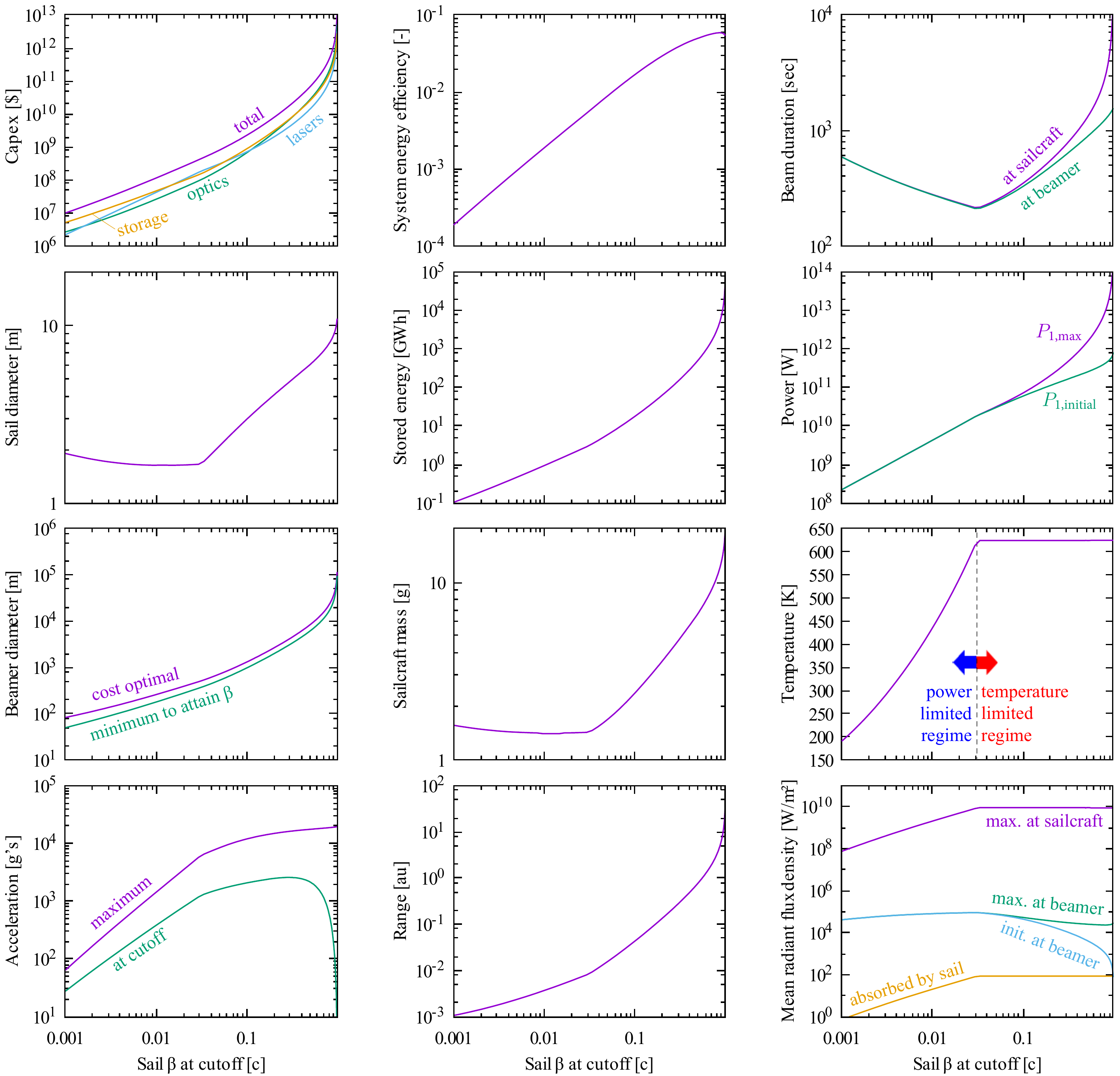}
\caption{Mission variation with cruise velocity}
\label{fig:02cmissioncruisevel}
\end{figure*}

\Cref{tab:02ctechvariations} summarizes point designs, each of which varies a technology figure of merit relative to the baseline.  The table lists the change in beamer capex, beamer diameter, beamer maximum transmit power, and sail diameter, relative to the baseline of \cref{tab:02coutputs}.  In the first entry, the laser cost per watt is increased by 10 times.  The resulting capex increases only by 2.4 times because the cost-optimum power $P_{1}$ halves and the cost-optimum beamer diameter $D_{b}$ increases (increases in transmission efficiency) to compensate.  Similarly, increasing the optics cost per unit area by 10 times causes the capex to increase by only 2.4 times.  In this case, the cost-optimum solution halves the beamer diameter and dramatically increases the power to compensate.  Increasing the storage \$/kWh by 10 times causes the capex to nearly quadruple, the biggest increase so far because storage is the dominant cost, as seen in the top left plot of \cref{fig:02cmissionsaildia}.  The cost-optimum point design has an increased system energy efficiency, achieved by increasing the product of beamer and sail diameters, as described by the Goubau beam propagation \cref{eq:goubautaudef,eq:goubauefficiency}.

In the happy event that a technology is even cheaper than needed, the capex is reduced less than might be intuitively expected.  Reducing laser cost to a tenth of its baseline has the least effect, reducing capex to 70\% of its baseline.  This is because, as seen in \cref{fig:02cmissionsaildia}, laser cost is subdominant and further reductions do not significantly affect the leading costs of storage and optics.  For both optics and storage, reducing their costs to a tenth of the baseline has the effect of halving the capex.

If the sail material absorbs 10 times more energy than expected\footnote{Reducing absorptance is a good way to reduce trajectory duration given that the sailcraft is mostly temperature-limited, as seen in \cref{fig:02ctrajectory}.  Shorter durations might be needed if the beam cannot be interrupted by passing satellites.  Absorbing 0.1x increases the temperature-limited irradiance on the sailcraft by 10x, acceleration increases by 10x, and the duration falls from \kpZeroTwoCDurationAccnMin\ to less than \SI{3}{\minute}.}, then the capex is more than quadruple that of the baseline, the largest increase of all. In comparison, a tenfold increase in payload mass does not quite triple the capex.  The three order of magnitude absorptance range considered here is smaller than the five order of magnitude absorptance range of the current sail substrate material candidates \cite{atwater2018materials}.  For this reason, \cref{tab:02ctechvariations} lists results for higher absorptances at the end.

In all, a tenfold increase in any figure of merit does not increase the capex by nearly as much.  Nor does a tenfold decrease in a technology cost decrease the capex by nearly as much.  This is because there are three constituent costs that are of comparable magnitudes after cost minimization.  The minimization process trades away large increases in one cost for smaller increases in others or all.  Therefore, cost minimization dampens the solution's sensitivity to unexpected changes in technology figures of merit.

Sail reflectance variations are also listed in \cref{tab:02ctechvariations}.  They show that if the baseline mission had used a single-layer dielectric sail with a reflectance of 25\%, as shown in \cref{fig:stratlayer}, then the capex would triple.  This \$18B difference is why the baseline mission uses a photonic crystal sail with a higher reflectance of 70\%.  It is well known that nanostructuring the sail can improve the reflectance \cite{atwater2018materials}, and with such a large financial return, it is inconceivable that this would not be done.  Though almost perfect reflectors are possible in principle, \cref{tab:02ctechvariations} shows diminishing financial returns much above 90\% reflectance.  Such performance is attainable at a single wavelength, but the trajectory-averaged value will likely be lower as the Doppler shift lengthens the beam wavelength by 20\% over the course of the sail acceleration.  Also, the sail material's optical properties may be leveraged for communications, sensing, or processing, and these competing uses would likely be worth a modest decrease in reflectance and/or increase in absorptance during sail acceleration.

\subsubsection{Variation with respect to cruise velocity}

Cruise velocity $\beta$ is specified to be \SI{0.2}{c} in the Breakthrough Starshot objectives.  Nevertheless, it is interesting to vary this velocity to see how the cost-optimal mission changes. The resulting plots are shown in \cref{fig:02cmissioncruisevel}. Each quantity is plotted from \SI{0.001}{c} to \SI{0.99}{c}. The beamer (total) capex is shown in the top left plot.  Also shown are constituent costs of storage, optics, and lasers.  Choosing a mission that has half the cruise velocity, \SI{0.1}{c}, decreases the beamer capex to a quarter of its baseline, whereas doubling it to \SI{0.4}{c} quintuples it.  That is to say, it costs \$29B extra to halve the trip time or saves \$6B to double it.  In the case of Alpha Centauri, this shortens the \SI{4.37}{ly} trip time from 22 years to 11 years, or lengthens it to 44 years.

\Cref{fig:02cmissioncruisevel} also shows that cost-optimal missions are limited by sail temperature if the cruise velocity is faster than \SI{0.03}{c}.  This value of cruise velocity corresponds to minimum beam duration.  The minimum beamer diameter reaches \SI{10}{\kilo\meter} at \SI{0.7}{c} and \SI{100}{\kilo\meter} at greater than \SI{0.99}{c}.

\subsection{\SI{0.01}{c} Precursor}
Precursor missions demonstrate the key technologies needed by Breakthrough Starshot, albeit at smaller scale, lower speed, and lower cost than the \SI{0.2}{c} missions.  The precursor mission point design embodies key elements of mid-21st century missions to probe the inner solar system through to the Oort cloud. A \SI{0.01}{c} cruise velocity enables the sailcraft to reach Mars in a day, Saturn in a week, the Kuiper belt in a month, or the minimum solar gravitational focus distance of \SI{550}{au} in a year. A successful precursor mission proves that technologies, people and processes are ready to scale up to the full \SI{0.2}{c} mission.
\subsubsection{Inputs}

The inputs to the precursor mission point design are summarized in \cref{tab:001cinputs}.  The \kpWavelength\ wavelength is consistent with ytterbium-doped fiber amplifiers.  An initial sailcraft displacement of \SI{300}{\kilo\meter} is consistent with a low Earth orbit from which the sailcraft is entrained by a low-power beam.  

\begin{table}[tbh]
\caption{System model inputs for \SI{0.01}{c} precursor mission}
\label{tab:001cinputs}
\centering
\begin{tabular}{l}
\hline			
\SI{0.01}{c} target speed \\
\kpWavelength\ wavelength \\
\SI{300}{\kilo\meter} initial sail displacement from laser source \\
\\
\SI{1}{\milli\gram} payload \\
\SI{0.2}{\gram\per\meter\squared} areal density \\
\kpAbsorptance\ spectral normal absorptance at \kpWavelength\\
40\% spectral normal reflectance at \kpWavelength\\
\kpMaxTemp\ maximum temperature\\
\kpEmittance\ total hemispherical emittance (2-sided, \kpMaxTemp) \\
\\
\SI[per-mode=symbol]{1}[\$]{\per\watt} laser cost \\
\$10k/m\textsuperscript{2} optics cost\\
\SI[per-mode=symbol]{100}[\$]{\per\KWH} storage cost\\ 
50\% wallplug to laser efficiency\\
70\% of beam power emerging from top of atmosphere\\
\hline  
\end{tabular}
\end{table}  

A \SI{1}{\milli\gram} payload (non-sail mass) is reserved for one or two sensors and associated support systems.  Sail mass is calculated by the system model based on the value of $D_{s}$ chosen by the optimizer combined with the areal density given as an input.  Similar to the \SI{0.2}{c} point design, this design assumes a photonic crystal sail material with the same thermal, mass and optical properties, except for a less ambitious reflectance of 40\%.  Again, the stratified layer optical model is turned off, and absorptance and reflectance remain constant throughout trajectory integration.

Cost factors are chosen to lie between present values and those of the \SI{0.2}{c} point design.  The laser cost is \SI[per-mode=symbol]{1}[\$]{\per\watt}, two orders of magnitude lower than the present cost, consistent with requirements for laser-powered launch vehicles, yet two orders of magnitude greater than microwave oven magnetrons.  The optics cost is \$10k/m\textsuperscript{2}, two orders of magnitude lower than the present cost for diffraction-limited optics, consistent with significant production line automation.  This value is comparable with the cost of radio telescope aperture.  \SI[per-mode=symbol]{100}[\$]{\per\KWH} energy storage is above current material cost floors for many energy storage technologies and in line with experience curve projections for electric vehicle battery packs circa 2035 \cite{schmidt2017future}.

\subsubsection{Results}
Upon running the system model using the inputs in \cref{tab:001cinputs}, the optimizers converge to the values given in \cref{tab:001coutputs}. The cost-optimum sail diameter is found to be \kpZZOCSailDiameter, corresponding to a mass of \kpZZOCSailMass.

\begin{table}[hbtp]
\caption{System model outputs for \SI{0.01}{c} mission}
\label{tab:001coutputs}
\centering
\begin{tabular}{l}
\hline			
\kpZZOCCapex\ beamer capex comprised of: \\
\qquad\kpZZOCLasers\ lasers (\kpZZOCPowerMax\ max. transmitted power)\\
\qquad\kpZZOCOptics\ optics (\kpZZOCBeamerDiameter\ primary effective diameter)\\
\qquad\kpZZOCStorage\ storage (\kpZZOCStored\ stored energy) \\
\\
\kpZZOCEnergy\ energy cost per mission (\kpZZOCStored\ @\SI[per-mode=symbol]{0.1}[\$]{\per\KWH}) \\
\kpZZOCSysEfficiency\ system energy efficiency\\
\\
\kpZZOCSailDiameter\ sail diameter\\
\kpZZOCSailMass\ sailcraft mass (includes payload mass)\\
\\
\kpZZOCDurationPulseMin\ (\kpZZOCDurationPulseSec) beam transmit duration\\
\kpZZOCDurationAccnMin\ (\kpZZOCDurationAccnSec) sail acceleration duration\\
\\
\kpZZOCPhotonPressure\ temperature-limited photon pressure\\
\kpZZOCPhotonForce\ temperature-limited force\\
\kpZZOCAccnInit\ temperature-limited acceleration\\
\kpZZOCAccnFin\ final acceleration (\kpZZOCDispFinAu), \kpZZOCDispFinLs\ from source\\
\\
\kpZZOCFluxBeamerMax\ beamer maximum beam radiant exitance\\
\kpZZOCFluxSailTLimited\ sailcraft temperature-limited irradiance\\
\hline
\end{tabular}
\end{table}

At \SI{0.01}{c}, the sailcraft has a relativistic kinetic energy of \kpZZOCRelKE, whereas the beamer uses \kpZZOCStored.  This yields \kpZZOCSysEfficiency\ system energy efficiency; two orders of magnitude lower than the \SI{0.2}{c} mission.  However, this energy costs only \kpZZOCEnergy\ at a price of \SI[per-mode=symbol]{0.1}[\$]{\per\KWH}, making it five orders of magnitude cheaper than the beamer capex of \kpZZOCCapex.

At \kpZZOCBeamerDiameter\ effective diameter, the \SI{0.01}{c} beamer is 16 times smaller than the \SI{0.2}{c} beamer.  Also, it has one third the maximum radiant exitance at \kpZZOCFluxBeamerMax\ vs. \kpZeroTwoCFluxBeamerMax.  The temperature-limited irradiance in the sailcraft frame is the same as that of the \SI{0.2}{c} mission because the sailcraft's absorptance, emittance, and temperature limit are the same.  The initial sailcraft acceleration is two thirds that of the \SI{0.2}{c} mission because of the lower reflectance, but the cost-optimal trajectory ends at only \kpZZOCAccnFin, squeezing almost everything it can from the diminishing beam.  Even then, the energy storage costs only \kpZZOCStorage\ because storage is cheap relative to lasers and optics, and because there is a limit to the extent that laser and optics costs can be reduced by increasing the pulse length.

The sailcraft's relativistic kinetic energy reaches \kpZZOCRelKEMJ\ (\kpZZOCRelKE).  Per unit mass, this is \kpZZOCRelKEspecific. In comparison, the heat produced by Pu-238 alpha decay is half as much at \SI{2}{\tera\joule\per\kilo\gram}.  At \SI{1}{AU}, the solar wind is primarily composed of \SI{9}{protons\per\centi\meter\cubed} flowing radially outward from the Sun at \SI{0.001}{c} \cite{schwenn2001solar}.  From the perspective of a sailcraft cruising away from the Sun, this manifests as a \SI{0.009}{c} proton beam that is incident from the direction of travel, having a combined kinetic energy of \SI{0.15}{\watt\per\meter\squared}, or only \SI{4.2}{\milli\watt} over the sail's area if it faces the direction of travel.

\subsection{Vacuum Tunnel}

\begin{figure*}[t]
\centering
\includegraphics[width=0.9\textwidth,keepaspectratio=true]{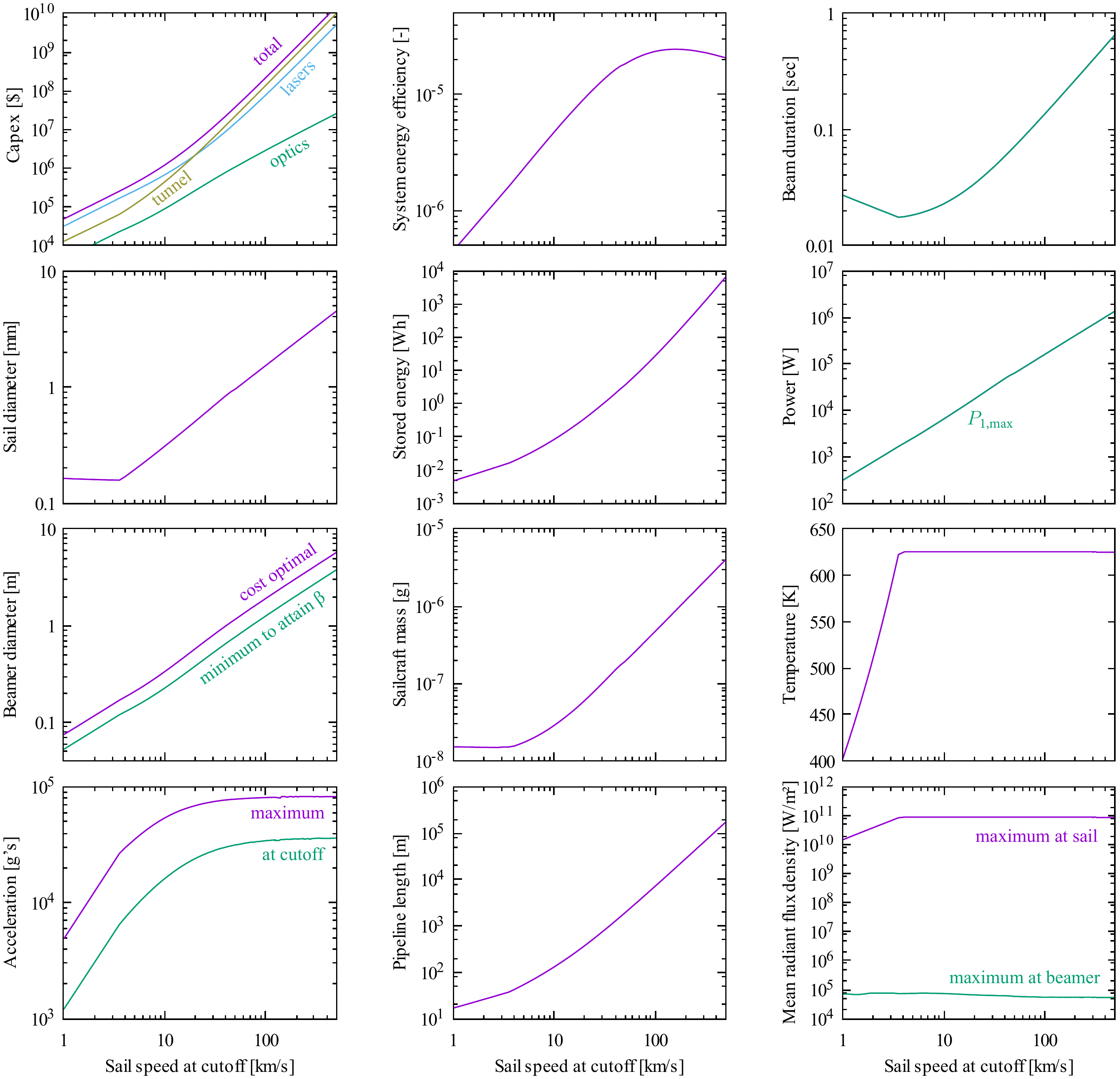}
\caption{Tunnel characteristics vs. sail speed at cutoff}
\label{fig:vactunnel}
\end{figure*}

Forward \cite{forward1986laser} recognized that sails can be levitated and tested using medium power lasers in the \SI{1}{g} gravity field of the Earth. Experimentally, this is a desirable configuration because the instrumentation stares at a stationary sail within a vacuum chamber.  After the sail beamrides at \SI{1}{g}, the dynamics then need to be tested and understood at ever-increasing accelerations.  This can be accomplished horizontally in a vacuum tunnel.  Gravity will try to pull the sail from the axis of the beam, and the sail should counteract this and other perturbations.  A key milestone is when the acceleration achieved by the sail equals the \kpZeroTwoCAccnInit\ acceleration needed by the \SI{0.2}{c} mission.  After that, the most visible milestone is reached when the sail becomes the fastest human-made craft.  The current record is held by the Helios-B probe, which achieved a heliocentric speed of \SI{70.22}{\kilo\meter\per\second} during its closest pass of the Sun in April 1976.  It is expected that this record will be broken by the \SI{200}{\kilo\meter\per\second} speeds of the upcoming Parker Solar Probe.  But is it practical for Breakthrough Starshot to attempt the speed milestone in a ground-based vacuum tunnel?

\subsubsection{Inputs}

\begin{table}[h]
\caption{System model inputs for the vacuum tunnel}
\label{tab:tunnelinputs}
\centering
\begin{tabular}{l}
\hline			
\kpWavelength\ wavelength \\
\SI{1}{\meter} initial sail displacement from laser source \\
\\
\SI{10}{\nano\gram} payload \\
\SI{0.25}{\gram\per\meter\squared} areal density \\
$10^{-9}$ spectral normal absorptance at \kpWavelength\\
35\% spectral normal reflectance at \kpWavelength\\
\kpMaxTemp\ maximum temperature\\
\kpEmittance\ total hemispherical emittance (2-sided, \kpMaxTemp)\\
\\
\SI[per-mode=symbol]{100}[\$]{\per\watt} laser cost \\
\kpTunnelInputOptics\ optics cost\\
\SI[per-mode=symbol]{500}[\$]{\per\KWH} storage cost\\
\kpTunnelInputTunnel\ vacuum tunnel cost\\  
50\% wallplug to laser efficiency\\
\hline  
\end{tabular}
\end{table}

The inputs to the vacuum tunnel system model are detailed in \cref{tab:tunnelinputs}.  Relative to the inputs for the \SI{0.01}{c} mission, the vacuum tunnel has a much shorter initial sail displacement of \SI{1}{\meter}, consistent with a sail positioned close to an optic at the start of the tunnel.  Also, the payload is reduced to a token \SI{10}{\nano\gram}, the mass of a few cells, and the reflectance is reduced to a near-term value of 35\% per the stratified layer reflectance predictions in \cref{fig:stratlayer}.  Such a reflectance is consistent with commercially-available Si$_{3}$N$_{4}$ membrane x-ray windows, for example.  Again, the stratified layer optical model is turned off, and the absorptance and reflectance remain constant throughout trajectory integration.

To equal the acceleration of the \SI{0.2}{c} mission, the sail ideally has the same areal density, reflectance, absorptance, and temperature limit as the \SI{0.2}{c} sail.  For the vacuum tunnel, point designs assume higher areal density and lower reflectance to be consistent with the idea that for the time being, a sail is a $\sim$ \SI{1}{\milli\meter} diameter high-purity film that is not a photonic crystal.  Correspondingly, the absorptance is assumed to be an order of magnitude lower than that of the \SI{0.2}{c} sail.  If the absorptance were not lower, then the reflectance would need to be increased to obtain sufficient acceleration.

The cost factors are chosen to be consistent with current market values.  In addition to laser, optics, and storage costs, there is an additional vacuum tunnel cost of \kpTunnelInputTunnel. This tunnel cost factor is consistent with the cost of the LIGO vacuum tunnels \cite{lindquist2002ligo} and includes the beam tunnel and its enclosure, as well as vacuum equipment. To account for differences from the \SI{1}{\meter} LIGO tunnel diameter, the tunnel cost is multiplied by the relative tunnel diameter. The tunnel cost does not include items that do not scale with length such as R\&D, detectors, project management, laboratory construction, and operations.

\subsubsection{Results}

Upon running the system model using the inputs in \cref{tab:tunnelinputs}, the model converges to the values plotted in \cref{fig:vactunnel}. This figure is comprised of a family of point designs spanning the range of 1-\SI{500}{\kilo\meter\per\second} sail speed at cutoff. Slower than \SI{20}{\kilo\meter\per\second}, laser cost is of primary importance and tunnel cost is secondary. Faster than \SI{30}{\kilo\meter\per\second}, tunnel cost is of primary importance and laser cost is secondary. In all cases, optics cost is tertiary. Storage cost turns out to be trivial, not even reaching \$1000 at \SI{500}{\kilo\meter\per\second}, so the cost plot is not scaled to show it.

If the sail had no payload, it would have a temperature-limited acceleration exceeding \SI{80000}{g's} regardless of tunnel length; this acceleration is more than five times that of the \SI{0.2}{c} mission.  When payload mass is taken into account, the temperature-limited acceleration exceeds that of the \SI{0.2}{c} mission only in tunnels longer than \SI{50}{\meter}.

For \SI{20}{\kilo\meter\per\second} sail speed at cutoff, sufficient to escape the solar system starting from Earth, the model infers:  A \kpTunnelAccnTLen\ tunnel, \kpTunnelPwr\ of lasers, and a \kpTunnelOpticsDia\ diameter telescope, costing a total of \kpTunnelAccnCCapex.  

For \SI{200}{\kilo\meter\per\second} sail speed at cutoff, which equals or exceeds the fastest human-made craft, the model infers: A \kpTunnelTwoAccnTLen\ tunnel, \kpTunnelTwoPwr\ of lasers, and a \kpTunnelTwoOpticsDia\ diameter telescope, costing a total of \kpTunnelTwoAccnCCapex.  

There is a small but noticeable bump toward the upper right corner of the sail diameter and other plots in \cref{fig:vactunnel}, near the \SI{50}{\kilo\meter\per\second} mark.  This bump is accompanied by a subtle change in gradient of sail diameter vs. speed.  Decreasing the integration step size and convergence tolerances does not remove the bump, nor does the model hit any kind of assumed limit on an internal variable.  Also, plotting costs vs. sail diameter for the speed at which the bump occurs shows that the minimum (found by iteration 1 of the solution procedure shown in \cref{fig:solutionproc}) is a global minimum and that there are no competing local minima that could cause the solution to jump from one value to another.  It turns out that the source of this bump is the Goubau beam model where the two efficiencies described by \cref{eq:goubauefficiency} are spliced together.  Though this splice preserves the continuity of the function, it introduces a discontinuity in its gradient.  Smoothing this function in the vicinity of the splice has the effect of removing the bump in \cref{fig:vactunnel}.

\section{Conclusions}

In this paper, a system model is formulated to describe a beam-driven sailcraft.  It minimizes beamer capital cost by trading off the relative expenses of lasers, optics, and storage.  The system model employs nested numerical optimizers and trajectory integration, whereas earlier models were based on closed-form approximations.  The outcome is that the solution is cheaper and generates more accurate requirements, but it also exhibits more complex behaviors. 

The system model is used to compute point designs for a \SI{0.2}{c} Alpha Centauri mission and a \SI{0.01}{c} solar system precursor mission.  Also, a family of solutions is computed for a ground-based vacuum tunnel in which beam-riding and other aspects of the sail can be tested.  All assume the case of a circular dielectric sail that is accelerated by photon pressure from a \kpWavelength\ wavelength beam. Earlier investigators were led astray by pursuing high reflectance alone, leading them to metallic sails and/or heavy multilayer dielectric sails.  This paper shows that reflectance divided by sailcraft areal density (or thickness, if the material is held constant) is a figure of merit that leads to substantially improved system performance relative to earlier sailcraft concepts.  A stratified layer model shows that the sail acceleration per unit power is maximized by a single-layer dielectric that is $\lambda/7$ thick.  An ideal technology would improve reflectance by removing, as opposed to adding, mass.  For this reason, two-dimensional nanohole photonic crystals are of interest as a future sail material.

The \SI{0.2}{c} point design minimizes capital cost by accelerating a \kpZeroTwoCSailDiameter\ diameter sailcraft for \kpZeroTwoCDurationAccnMin.  In minimizing the cost, it is surprising that laser costs are secondary to optics and storage costs, because this implies that there is a limit to the amount that cheap laser power can compensate for expensive optics or storage.  The point design assumes \SI[per-mode=symbol]{0.01}[\$]{\per\watt} lasers and \SI[per-mode=symbol]{500}[\$]{\per\meter\squared} optics to achieve \kpZeroTwoCCapex\ capital cost for the ground-based beamer. In contrast, the energy needed to accelerate each sailcraft is a thousand times cheaper, making the \kpZeroTwoCSysEfficiency\ system energy efficiency unimportant. With large fixed costs and low incremental costs, why not use the beamer to propel sailcraft to every reachable star as often as possible?  Therefore, a rational outcome of Breakthrough Starshot might be to pave the way for multi-lightyear pipelines of sailcraft that fly past each target star every few weeks \textit{ad infinitum}. For the \SI{0.2}{c} mission, it is clear that the sailcraft must be beam riding; the speed of light is too slow to allow the beam to follow the sailcraft.  If the beamer slews the beam during acceleration, it will be to dodge satellites or fine-tune the sailcraft's destination.  It is also clear that nuclear batteries are dead weight, having a thousand times lower specific energy than the sailcraft's kinetic energy.  From the sailcraft's perspective cruising at \SI{0.2}{c}, the interstellar medium manifests as a \SI{0.7}{\kilo\watt} monoenergetic hydrogen beam that is incident from the direction of travel.  A key question for future research is, what fraction of this power can be harvested?

For an extra \$29B, the cruise velocity can be doubled to \SI{0.4}{c}, which halves the cruise time to Alpha Centauri to 11 years. Following a successful \SI{0.2}{c} mission, there would be high confidence in incrementally upgrading the beamer to support \SI{0.4}{c} missions.  Hence, it is not unreasonable to expect that private and government investments over several years could amount to \$29B or more.  However, with increasing cruise velocity comes increasing beamer diameter.  There is a practical and desired limit to beamer diameter, but how large is it? Cities are perhaps the best guide because they are human-engineered surfaces of the largest diameter.  Greater London has a city area of \SI{1572}{\kilo\meter\squared}.  If it were circular, it would have a diameter of \SI{45}{\kilo\meter}. \Cref{fig:02cmissioncruisevel} predicts that a beamer the size of London would be capable of propelling a sailcraft to greater than \SI{0.9}{c}.

The \SI{0.01}{c} point design differs from the \SI{0.2}{c} point design in that it assumes nearer-term cost factors of \SI[per-mode=symbol]{1}[\$]{\per\watt} and \$10k/m\textsuperscript{2} to achieve \kpZZOCCapex\ capital cost for its beamer and \kpZZOCEnergy\ energy cost per \kpZZOCSailDiameter\ diameter sailcraft. This sailcraft is 20 times smaller and 600 times lighter than the \SI{0.2}{c} sailcraft, so its various subsystems need to be integrated into a much smaller mass and area.  But, for such a small photonic crystal sailcraft, what functionality is theoretically possible and what sail area will it take?  The \SI{0.2}{c} mission will be bombarded by dust and radiation over two decades, so it may need many duplicates of each subsystem to reach its mission objective.  Hence, the \SI{0.01}{c}  sailcraft could be developed as a single unit cell of the \SI{0.2}{c} sailcraft.  If low mass and low energy cost translate into low incremental costs, then it makes sense to use many precursors to prove the sailcraft technologies.  Precursors can probe magnetic fields and dust particle fluxes ahead of the \SI{0.2}{c} missions.  Indeed, the whole heliosphere is a testbed for sailcraft communication and sensor technologies. At \SI{0.01}{c}, the Kuiper belt is only a month away.  Unlike the \SI{0.2}{c} sailcraft, the \SI{0.01}{c} sailcraft has very little power:  Its specific kinetic energy is comparable with the specific heat of Pu-238 alpha decay.  From the perspective of a sailcraft cruising at \SI{0.01}{c}, the solar wind manifests as only a \SI{4}{\milli\watt} proton beam incident from the direction of travel.  In comparison, the solar power available at Kuiper belt distances of \SI{30}{au} is an order of magnitude higher. With solar power only, inner solar system missions are of course easier.

The ground-based vacuum tunnel assumes present-day cost factors of \SI[per-mode=symbol]{100}[\$]{\per\watt} lasers, \kpTunnelInputOptics\ optics, and \$10k/m vacuum tunnel for its family of point designs spanning the range of 1-\SI{500}{\kilo\meter\per\second} maximum sail speed.  Primarily, a tunnel would exist to test sail beam-riding dynamics.  Secondarily, the tunnel could be used to demonstrate high sail velocities, and doing so would prove mastery of beam-riding dynamics transverse to the optical axis and also longitudinally as the beam focus accelerates. Under the material properties assumed here, speed milestones exceeding \SI{100}{\kilo\meter\per\second} might be achieved at lowest cost by using space-based missions as opposed to ground-based tunnels.  However, tunnels may still be cheapest if the incremental cost of upgrading an existing facility is low enough.  Also, tunnel length and cost depend on sail material properties, so this conclusion should be revisited as sail materials become better characterized.

A key thesis of Starshot is that the cost of lasers and optics can and will fall.  For \SI{0.2}{c} missions, the laser cost per unit power and optics cost per unit area are chosen such that the beamer costs less than \$10B.  Consequently, \SI[per-mode=symbol]{0.01}[\$]{\per\watt} laser cost and \SI[per-mode=symbol]{500}[\$]{\per\meter\squared} optics cost should be understood as requirements that derive from the \$10B cost cap.  These requirements are to be achieved through technology development and production line automation.  Unconstrained by a cost cap, nearer-term precursor missions assume less favorable values for laser and optics costs, and the beam tunnel calculations assume present-day values.  Based in part on requirements derived from the system model, the array elements that comprise the beamer will become better defined in future.  Once designs are available for the beamer array elements, it makes sense to update the cost inputs for the \SI{0.2}{c} mission based on calculated laser and optics material cost floors combined with an analysis of the array element production chain.

The results presented here are based on uncertain input values, and \cref{tab:02ctechvariations} shows that the cost-optimal solutions are surprisingly resilient to them.  Of the inputs, laser and optics costs are the most uncertain.  Their orders of magnitude uncertainty can be reduced by studies of materials cost floors and production automation.  The sail material is also an important source of uncertainty, in particular its absorptance, and to lesser extents its emittance, reflectance, areal density and maximum temperature.  Candidate sail materials are wide band-gap dielectrics because their absorptances are very low and rise the least as temperature rises.  This dependence of absorptance on temperature is important and needs to be captured in future versions of the system model. Because of the large number of sail design parameters to explore, near-term work should define perhaps three promising reference materials and characterize their temperature-dependent properties and other properties as needed by the system model.  Given current uncertainties, \cref{tab:02ctechvariations} shows that the sailcraft material properties make more than a billion-dollar difference to the beamer cost.  Hence, it is worth considerable R\&D investment to characterize and optimize materials specifically for laser-driven sailing.

Finally, the modeling work presented here takes place within the context of a wider systems engineering effort.  In future, system model development will be driven by the increasing need to infer requirements and margins from an up-to-date and self-consistent understanding of the problem.  The system model can be extended to handle probabilistic inputs in order to estimate parameter sensitivities, and to bound requirements and margins. It can also be extended to incorporate domain models. Such models usually add computational complexity, so they need to be blended with the existing simplistic models to converge system point designs in an acceptable time frame.  Domain models would add greater detail and accuracy in the areas of beamer elements and phasing; transatmospheric beam propagation; communications; cost modeling; and sailcraft thermal, structural, and electromagnetic properties.

\section{Acknowledgments}
This work was supported by the Breakthrough Prize Foundation. I acknowledge helpful discussions with Pete Klupar, Bruce Draine, James Benford, Ognjen Ilic, and Harry Atwater.
\section*{References}
\bibliography{K:/Databases/Bibtex/kp}
\end{document}